\documentstyle[11pt,amsfonts]{article}

\newcommand{\be}{\begin{equation}}
\newcommand{\ee}{\end{equation}}
\newcommand{\bea}{\begin{array}}
\newcommand{\ea}{\end{array}}
\newcommand{\beqa}{\begin{eqnarray}}
\newcommand{\eeqa}{\end{eqnarray}}
\newcommand{\bean}{\begin{eqnarray*}}
\newcommand{\eean}{\end{eqnarray*}}

\def\BI{{\rm 1\!l}}

\def\up#1{\leavevmode \raise.16ex\hbox{#1}}

\newcommand{\gapproxeq}{\lower
 .7ex\hbox{$\;\stackrel{\textstyle >}{\sim}\;$}}
\newcommand{\lapproxeq}{\lower .7ex\hbox{$\;\stackrel
{\textstyle <}{\sim}\;$}}

\newcounter{appendice}

\def\thebibliography#1{{\bf REFERENCES\markboth
 {REFERENCES}{REFERENCES}}\list
 {[\arabic{enumi}]}{\settowidth\labelwidth{[#1]}\leftmargin\labelwidth
 \advance\leftmargin\labelsep
 \usecounter{enumi}}
 \def\newblock{\hskip .11em plus .33em minus -.07em}
 \sloppy
 \sfcode`\.=1000\relax}

\begin{document}

\centerline{ \LARGE   Particle Classification and Dynamics }
\medskip
\centerline{ \LARGE   in $GL(2,C)$ Gravity}
\vskip 2cm

\centerline{ {\sc  A. Stern }  }

\vskip 1cm

\centerline{  Dept. of Physics and Astronomy, Univ. of Alabama,
Tuscaloosa, Al 35487, U.S.A.}

\vskip 2cm

\vspace*{5mm}

\normalsize
\centerline{\bf ABSTRACT}

A relatively simple approach to noncommutative gravity utilizes the gauge theory formulation of general relativity and involves replacing the Lorentz gauge group  
by a larger group. This results in additional field degrees of freedom which either must  be constrained to vanish in a nontrivial way, or require physical interpretation.  With the  latter in mind, we examine the coupling of the  additional fields to point particles.   Nonstandard particle degrees of freedom should  be introduced in order to write down the most general coupling. The example we study is the $GL(2,C)$ central extension of  gravity given by  Chamseddine, which  contains  two  $U(1)$ gauge fields, and a   complex vierbein matrix,  along with the usual spin connections.  For the general coupling one should then attach  two $U(1)$ charges and a  complex momentum vector to the particle, along with the spin.  The momenta span orbits in a   four-dimensional  complex vector space, and are classified by   $GL(2,C)$ invariants and by their little groups.  In addition to orbits associated with  standard massive and massless particles, a number of novel orbits can be identified.  We write down a general action principle for particles associated with any nontrivial orbit and show that it leads to corrections to geodesic motion.
We also examine the classical and quantum theory of the particle in flat space-time.

\vskip 2cm
\vspace*{5mm}

\newpage
\scrollmode
\section{Introduction}

The standard gauge theory formalism for gravity  \cite{Utiyama:1956sy},\cite{Kibble:1961ba},\cite{Hehl:1976kj},\cite{Chamseddine:2005td} is based on the Lorentz group, or equivalently on its $SL(2,C)$ covering group.  A central extension  to the $GL(2,C)$ gauge group has been proposed  by Chamseddine and some  properties of the resulting  theory have been investigated.\cite{Chamseddine:2003we}
The $GL(2,C)$ gauge theory  has the advantage over the standard gauge theory formulation in that it allows for a straightforward generalization to the noncommutative version of the theory.  Unlike in the more sophisticated treatment of noncommutative gravity given by Aschieri et. al. \cite{Aschieri:2005yw}, the diffeomorphism symmetry of the commutative theory is not preserved in this approach.  However, its technical simplicity makes it much more  amenable for practical applications.  These applications include the computation of noncommutative corrections to the known solutions of general relativity.  Such computations have been of recent interest.\cite{Pinzul:2005ta},\cite{Balachandran:2006qg},\cite{Dolan:2006hv},
\cite{Chaichian:2007we},\cite{Mukherjee:2007fa},\cite{Kobakhidze:2007jn},\cite{Banerjee:2008gc},\cite{Buric:2008th},\cite{Nicolini:2008aj},\cite{Fabi:2008ta}. 

Although the noncommutative  generalization of  $GL(2,C)$ gauge theory is straightforward, its physical interpretation as a gravity theory is not - due to the presence of additional field degrees of freedom. 
Two different interpretations of the noncommutative $GL(2,C)$ gauge theory are possible, which we now mention:
\begin{enumerate}

\item  One approach is to eliminate the additional degrees of freedom by expressing the  noncommutative $GL(2,C)$ gauge fields in terms of the commutative $SL(2,C)$ gauge fields using the Seiberg-Witten map\cite{Seiberg:1999vs}.  The standard  metric tensor, Lorentz curvature and torsion of the commutative theory can be utilized to determine the physical consequences of  the noncommutative dynamics. The disadvantage of this approach is that Seiberg-Witten map leads to complicated nonlinear and nonlocal constraints which must then be imposed on the noncommutative fields.\footnote{In a very recent article \cite{Aschieri:2009ky},  conditions involving just the charge conjugation operator are imposed on the noncommutative fields which eliminate the additional degrees of freedom in the noncommutative limit.  However, they do not eliminate  the additional degrees of freedom from the noncommutative theory without also using the Seiberg-Witen map.}  
   Solving the field equations with these constraints would  be a formidable task. (Actually, just writing down the field equations  is nontrivial in this case, because the noncommutative action should be varied with respect to the independent fields of the commutative theory.) 
 
\item  One drops these complicated constraints in the second approach, and instead treats all the $GL(2,C)$ gauge fields as independent degrees of freedom.  This  makes the field equations easier to solve, but  has  the disadvantage of introducing fields in the gravitational theory which have no analog in the  standard gauge theory formulation.  The degrees of freedom now include two $U(1)$ gauge fields,  a set of  complex  vierbein fields  and the usual spin connections. A physical interpretation of the extra degrees of freedom is then required in this approach and this can  already be addressed in the commutative  theory.  
\end{enumerate}
  
  Motivated by the second approach, we shall examine the physical content of the  $GL(2,C)$ central extension of the standard gauge theory formulation of  gravity by  coupling to test particles.   We require that the particle interaction be invariant under $GL(2,C)$ gauge transformations, as well as general coordinate transformations and reparametrizations of the evolution parameter.   In this regard, it is easy to find  a $GL(2,C)$ invariant metric tensor, from which the usual point particle Lagrangian can be constructed.  Geodesic motion with respect to the $GL(2,C)$ invariant metric tensor results. However, this particle Lagrangian is not general because it does not take into account all possible particle degrees of freedom.  These include the particle spin and two $U(1)$ charges, the latter of which can couple to the two $U(1)$ gauge fields in  the  $GL(2,C)$ gauge theory. 
  
   More curious is the fact that a  complex momentum vector, or equivalently two real momentum vectors, should be attached to the particle in order to couple to the complex vierbein fields mentioned above. Under the action  of $GL(2,C)$, these momentum vectors span orbits in a four-dimensional complex  vector space.  In analogy with the usual classification of  orbits in ${\mathbb{R}}^4$  for relativistic  particles, here particles are  classified by  orbits in a  ${\mathbb{C}}^4$ (or equivalently, ${\mathbb{R}}^8$).    The latter are labeled by  
 $GL(2,C)$ invariants, or  by their little groups. One of the $GL(2,C)$ invariants is quadratic and can be associated with the `mass', while another is quartic and does not have a familiar interpretation. One additional invariant can also  be found.  Up to now the discussion has not taken into account the particle spin (or charges).  Three  more invariants can easily be constructed (from the analog of Pauli-Lubanski vectors) when spin is present.  Due to the large number of invariants, a  large variety of different classes of orbits are possible. One such class of orbits  can be identified with standard massive particles, while several other disconnected orbits can be used to describe massless particles. Some nonstandard  orbits can be identified as well.
 
   An action principle can be formulated which is applicable to all of the nontrivial orbits, and it is a generalization of the action for a relativistic spinning particle.\cite{Balachandran:1979ha} (See also \cite{Horvathy:1977ne}.)
   The particle action is constructed from the real invariant bilinears for  $GL(2,C)$.   In addition to  being invariant under  $GL(2,C)$ gauge transformations, it is also invariant under general  coordinate transformations, reparametrizations in the evolution parameter and transformations generated  by the orbit's little group.  Coupling of the spin and the two $U(1)$ charges is achieved with the use of a Wess-Zumino type term, and the corresponding equations of motion are a generalization of the  Mathisson-Papapetrou equations\cite{Math},\cite{Papapetrou:1951pa} to the $GL(2,C)$ gauge theory.  They contain the Lorentz forces associated with the two $U(1)$ fields.
   A general class of  solutions to the equations of motion can be found.  We  show that they lead to deviations from geodesic motion  even in the case of  zero spin and charge.  The usual dynamics for relativistic particles is recovered upon specializing to flat space-time, although for one class of   orbits studied here, the particle can contain additional degrees of freedom. 
  
This article is organized as follows:  In section two we review the standard  gauge theory formulation of gravity based on the $SL(2,C)$ gauge group.  The extension to  $GL(2,C)$ gauge group is given in section three.  There we write down the $GL(2,C)$ invariant metric tensor, as well as other invariants of the theory, and the standard  particle Lagrangian  obtained from that metric tensor is presented. A  classification of particles based on orbits in the four-dimensional complex  momentum space is given in section four.
 We write down  a  $GL(2,C)$ invariant action  for  arbitrary orbits  in section five and  obtain the Euler-Lagrange equations of motion along with the general  solutions.   With the inclusion of  the  $GL(2,C)$ invariant Wess-Zumino term, the action is generalized to include interactions with the particle spin and two $U(1)$ charges in section six.   We specialize to flat space-time in section seven where the   $GL(2,C)$ gauge symmetry is broken.  In the quantum theory, the particle carries representations of a $16-$dimensional algebra, containing the Poincar\'e algebra.  We write down the algebra in section eight and construct the Hilbert space using the method of induced representations.  Concluding remarks are given in section nine.

\section{Standard gauge theory formulation  of gravity} 

\setcounter{equation}{0}

In the standard gauge theory formulation of gravity\cite{Utiyama:1956sy},\cite{Kibble:1961ba},\cite{Hehl:1976kj},\cite{Chamseddine:2005td}, one introduces spin connection and vierbeins,  $\omega^{ab}_\mu=-\omega^{ba}_\mu$ and $e^a_\mu$,  respectively.    $a,b,...=0,1,2,3$ are Lorentz indices which are raised and lowered using the flat metric tensor $[\eta_{ab}]={\rm diag} (-1,1,1,1)$, and $\mu,\nu,\cdots$ denote the space-time indices.  The space-time metric is
\be g_{\mu\nu}=e^a_{\mu}e^b_{\nu}\eta_{ab} \;,\label{mtrcnendom} \ee
and it is invariant under local Lorentz transformations.  Infinitesimal Lorentz variations $\delta_\lambda$ of  $\omega^{ab}_\mu$ and $e^a_\mu$ 
are of the form
\beqa \delta_\lambda \omega^{ab}_\mu &=& \partial_\mu \lambda^{ab} +  \omega^{ac}_\mu\lambda_c^{\;\;b} -  \omega^{bc}_\mu\lambda_c^{\;\;a} \cr & & \cr \delta_\lambda e^a_\mu &=&  e^b_\mu \lambda_b^{\;\;a}\label{slcvrtns}\;,\eeqa   where   $\lambda^{ab}=-\lambda^{ba}$ are infinitesimal gauge parameters.
 The Lorentz curvature  and torsion are defined by
 \beqa R^{ab}_{\mu\nu}&=&\partial_{[\mu} \omega^{ab}_{\nu]}  + \omega_{[\mu}^{ac}\omega_{\nu ] c}^{\;\;\;\;\;b}\;\label{lrnzcrvtr}\\ & &\cr  \tau^a_{\mu\nu}&=&\partial_{[\mu} e^{a}_{\nu]}  + \omega^{ab}_{[\mu} e_{\nu] b}
\;, \label{sltctrsn}\eeqa respectively, and satisfy the Bianchi identities
\beqa \partial_{[\mu } R^{ab}_{\nu\rho]} - R^a_{\;\;c[\mu\nu} \omega^{cb}_{\rho ]} + \omega^a_{\;\;c[\mu} R^{cb}_{\nu\rho]} &=&0  \cr & &\cr \partial_{[\mu } \tau^{a}_{\nu\rho]} - R^a_{\;\;b[\mu\nu} e^{b}_{\rho ]} + \omega^a_{\;\;b[\mu} \tau^{b}_{\nu\rho]} &=&0  
\eeqa  The Lagrangian density for pure gravity is proportional to the Lorentz invariant 
\be \epsilon_{abcd}\epsilon^{\mu\nu\rho\sigma}R^{ab}_{\mu\nu} e^c_\rho e^d_\sigma \ee

It shall be  convenient for us to introduce $\gamma-$matrices and utilize Dirac spinor notation.\cite{Isham:1972br} We  define 
 \be \omega_\mu =\frac 12 \omega^{ab}_\mu\sigma_{ab} \qquad \qquad e_\mu = e^a_\mu \gamma_a  \;,\ee
with $\{\gamma_a,\gamma_b\}=2\eta_{ab}\BI$   and $\sigma_{ab}= -\frac i4 [\gamma_a,\gamma_b]$. $\BI$ denotes the $4\times 4$ unit matrix and $[\;,\;]$ the matrix commutator.
Then for example, (\ref{mtrcnendom}) can be written as
\be g_{\mu\nu}= \frac 14{\rm tr}\;e_{\mu}e_{\nu}\;,\ee  using  ${\rm tr}\; \gamma_{a}\gamma_b = 4 \eta_{ab}$, while  (\ref{slcvrtns}) becomes  
\beqa \delta_\lambda \omega_\mu &=& \partial_\mu\lambda +i[\omega_\mu,\lambda] \cr & &\cr \delta_\lambda e_\mu &=& i[e_\mu,\lambda]\;, \label{lcllrnz} \eeqa
 where $\lambda=\frac 12 \lambda^{ab} \sigma_{ab}$.
 
 \section{Extension to $GL(2,C)$ gauge theory}
 
 \setcounter{equation}{0}
 
 \subsection{Motivation}

The Lorentz [or $SL(2,C)$] algebra no longer closes upon going to the noncommutative version of the standard gauge theory formulation.
  In the canonical approach to noncommutative field theories, one replaces the point wise product between functions by the star product, more specifically,  the Groenewold-Moyal star product
\be \star = \exp\;\biggl\{ \frac {i}2 \Theta^{\mu\nu}\overleftarrow{
  \partial_\mu}\;\overrightarrow{ \partial_\nu} \biggr\} \;\label{gmstr} \ee  Here
 $\Theta^{\mu\nu}=-\Theta^{\nu\mu}$ are constant matrix elements corresponding to the noncommutativity parameters and
  $\overleftarrow{
  \partial_\mu}$ and $\overrightarrow{ \partial_\mu}$ are,
respectively, left and right
derivatives with respect to  some coordinates $x^\mu$ of a smooth manifold. The commutator $[A,B]$  between any two matrix-valued  functions $A$ and $B$ in the commutative theory is then replaced by the star-commutator,
$[ A, B]_\star = A\star B -  B\star A$ in the noncommutative theory.  As a consequence, the commutators $[\omega_\mu,\lambda]$ and $ [ e_\mu,\lambda]$ appearing in the  gauge variation (\ref{lcllrnz}) are replaced by
\beqa [\omega_\mu,\lambda]_\star &= &\frac 18 \{ \omega^{ab}_\mu,  \lambda^{cd}\}_\star\; [\sigma_{ab},\sigma_{cd}] + \frac 18 [ \omega^{ab}_\mu,  \lambda^{cd}]_\star\; \{\sigma_{ab},\sigma_{cd}\}\label{clsron}\\ & &\cr
 [ e_\mu,\lambda]_\star &=& \frac 14 \{ e^{a}_\mu,  \lambda^{bc}\}_\star\; [\gamma_{a},\sigma_{bc}] + \frac 14 [ e^{a}_\mu,  \lambda^{bc}]_\star\; \{\gamma_{a},\sigma_{bc}\} \;\label{clsrtw}\eeqa
 in the noncommutative theory.
Here $\{\;,\;\}$ denotes the anticommutator, and  $\{\;,\;\}_\star$  the star-anticommutator,
$\{ a,b\}_\star = a\star b +  b\star a.$  For the Groenewald-Moyal star, $[ a, b]_\star$ is imaginary for any two real-valued functions $a$ and $b$, while $\{ a, b\}_\star$ is real.   The $SL(2,C)$ gauge algebra no longer closes due to the presence of the second term on the right hand side of  (\ref{clsron}).  Moreover, from the right hand side of (\ref{clsrtw}), noncommutative gauge transformations do not leave the space of vierbeins invariant.  The anticommutator  $\{\sigma_{ab},\sigma_{cd}\}$ appearing in (\ref{clsron}) is a linear combination of $\gamma_5$ and the unit matrix $\BI$, while the anticommutator $ \{\sigma_{ab},\gamma_c\}  $ appearing in (\ref{clsrtw}) is a linear combination of $\gamma_5\gamma_c$.

 Following 
 \cite{Chamseddine:2003we},  closure of the gauge algebra is recovered upon enlarging the gauge group from $SL(2,C)$ to $GL(2,C)$. For this one introduces $GL(2,C)$ connections $ A_\mu$ and  infinitesimal gauge parameters $\Lambda$ 
\be  A_\mu = \omega_\mu+  a_\mu\BI+ i b_\mu\gamma_5\qquad \qquad \Lambda = \lambda+ \alpha\BI + i \beta \gamma_5\;, \label{ALmbd}\ee where $ a_\mu$ and $ b_\mu$ are two $U(1)$ potentials and $ \alpha$ and $ \beta$ are two infinitesimal functions on space-time.  In addition, ref. \cite{Chamseddine:2003we} replaces $e^a_\mu$ with a  complex vierbein matrix $e^a_\mu+if^a_\mu $. Equivalently, upon writing
\be   E_\mu=  e_\mu +f_\mu\;,\qquad \quad f_\mu= f^a_\mu \gamma_5\gamma_a  \;, \label{gl2cvrbns}\ee
one can then write down a consistent set of
 noncommutative $GL(2,C)$ gauge variations $\delta_\Lambda$
\beqa \delta_\Lambda  A_\mu &=& \partial_\mu \Lambda +i[ A_\mu, \Lambda]_\star \cr & &\cr  \delta_\Lambda  E_\mu &=& i[ E_\mu,\Lambda]_\star\; \label{ncvlcllrnz} \eeqa  This leads to rather involved variations for the component fields $\omega^{ab}_\mu $, $a_\mu$, $ b_\mu$, $e^a_\mu$ and $ f^a_\mu$. \cite{Chamseddine:2003we}  

As stated in the introduction, our interest is to study the physical content of the new degrees of freedom in this model; i.e., those not present in the $SL(2,C)$ gauge theory formulation of gravity. They are contained in the fields  $a_\mu$, $b_\mu$ and $f^a_\mu$.  Thanks to the existence of the Seiberg-Witten map\cite{Seiberg:1999vs} between  commutative and noncommutative gauge theories, these issues can be addressed at the commutative level, meaning $\Theta^{\mu\nu}=0$.  The $GL(2,C)$ gauge  variations (\ref{ncvlcllrnz})
reduces to
\beqa \delta_\Lambda  A_\mu &=& \partial_\mu \Lambda +i[ A_\mu, \Lambda] \cr & &\cr  \delta_\Lambda  E_\mu &=& i[ E_\mu,\Lambda] \label{gl2cgtrns} \eeqa in this limit, and the resulting  variations of the component fields are now easy to write down:
\beqa \delta_\Lambda \omega^{ab}_\mu &=& \partial_\mu \lambda^{ab} +  \omega^{ac}_\mu\lambda_c^{\;\;b} -  \omega^{bc}_\mu\lambda_c^{\;\;a} \\ & &\cr \delta_\Lambda a_\mu &=& \partial_\mu \alpha \label{uoneltlgrp}\\ & &\cr \delta_\Lambda b_\mu &=& \partial_\mu \beta \label{brknuone}\\ & &\cr \delta_\Lambda e^a_\mu &=&  e^b_\mu \lambda_b^{\;\;a}         +2 f^a_\mu \beta\label{emuavrtn}\\ & &\cr \delta_\Lambda f^a_\mu &=&  f^b_\mu \lambda_b^{\;\;a}          +2 e^a_\mu \beta\;
\label{cmpsgtrns}\eeqa  
The two sets of veirbeins are invariant under the action of one of the $U(1)$ subgroups of $GL(2,C)$, while the get mixed under the action of the other $U(1)$.
A $GL(2,C)$ invariant field action was found in \cite{Chamseddine:2003we}, which  in the linear approximation yielded   massive modes in addition to the massless graviton.  

In what follows we shall introduce a test particle in the commutative $GL(2,C)$ gauge theory and examine possible $GL(2,C)$ invariant interactions.  We therefore need to construct  $GL(2,C)$ invariants, one of which should be the metric tensor.

\subsection{The metric tensor and other $GL(2,C)$ invariants}

We  need to identify a metric tensor for the $GL(2,C)$ gauge theory in order to connect it to a theory of space-time.  We require that the metric tensor transform nontrivially only under general coordinate transformations.  It should therefore  be invariant under the action of the $GL(2,C)$ gauge group.   We note in this regard that  (\ref{mtrcnendom}) is only invariant under the $SL(2,C)$ subgroup of $GL(2,C)$ and can no longer serve as the metric tensor.  In order to recover the $SL(2,C)$ gauge theory  when $f^a_\mu\rightarrow 0$, we need that the metric tensor reduce to (\ref{mtrcnendom}) in this limit.

 Two space-time dependent $GL(2,C)$ invariant bilinears can be constructed from the two sets of vierbeins in $E_\mu$:
\beqa {\tt g}_{\mu\nu} &=& \frac 14 {\rm tr}\; E_\mu E_\nu =e^a_{\mu}e_{a\nu} -f^a_{\mu}f_{a\nu}\label{gltcmtrc}\\& &\cr {\tt B}_{\mu\nu} &=& \frac 14 {\rm tr}\;\gamma_5 E_\mu E_\nu = f^a_{\mu}e_{a\nu} -e^a_{\mu}f_{a\nu}\;,\label{ntsmtrcblnr}\eeqa
where $\gamma_5=i \gamma_0\gamma_1\gamma_2\gamma_3 =-i\gamma^0\gamma^1\gamma^2\gamma^3$ and we used $ {\rm tr}\;\gamma_5 \gamma_{a}\gamma_b =0 $. 
Higher order invariants can also be defined; e.g.,
\beqa {\tt h}_{\mu\nu\rho\sigma}&=& {\rm tr}\; E_\mu E_\nu E_\rho E_\sigma\;\cr & &\cr  {\tt k}_{\mu\nu\rho\sigma}&=& {\rm tr}\;\gamma_5 E_\mu E_\nu E_\rho E_\sigma \label{honvrnts}\eeqa
The quadratic invariant ${\tt g}_{\mu\nu}$ given in (\ref{gltcmtrc}) is symmetric in the space-time indices and it reduces to (\ref{mtrcnendom}) when $f^a_{\mu}$ vanish.  It  can therefore be identified with the metric tensor in the $GL(2,C)$ gauge theory.  ${\tt B}_{\mu\nu}$ and ${\tt k}_{\mu\nu\rho\sigma}$ are antisymmetric 
 in all space-time indices, while  ${\tt h}_{\mu\nu\rho\sigma}$ is symmetric under cyclic permutations.  The volume integral of ${\tt k}_{\mu\nu\rho\sigma}$ serves as a cosmological term in the gravity action.\cite{Chamseddine:2003we} ${\tt B}_{\mu\nu}$, as well as   ${\tt g}_{\mu\nu}$, can be used to write down $GL(2,C)$ invariant couplings  to strings.  Here, however,  we shall only be concerned with point particles. 

The flat space-time metric tensor is recovered for  $E_\mu$ equal to 
 \be E^{flat}_\mu= c_1\gamma_\mu +c_2\gamma_5\gamma_\mu \;,\label{fltLmu}\ee
 where  the constants $c_1$ and $c_2$ satisfy  $c_1^2-c_2^2=1$.  The vacuum (\ref{fltLmu}) breaks the $GL(2,C)$ gauge symmetry to a $U(1)$ gauge symmetry, being associated with the variations (\ref{uoneltlgrp}), in addition to a global Lorentz symmetry.
  Massive and  massless modes were shown to follow from a $GL(2,C)$ invariant field action upon expanding  about the   flat space-time metric (\ref{fltLmu})
\be E_\mu=E^{flat}_\mu +  \bar e^a_\mu \gamma_a+\bar f_\mu^a\gamma_5\gamma_a\;,\label{grvtnmds}\ee
where   $ \bar e^a_\mu $ and $\bar f_\mu^a$ are small perturbations.\cite{Chamseddine:2003we}   The massless modes were shown to have spin two and were thus  identified with gravitons. 
  They correspond to the linear combinations
$\rho^a_\mu = c_1 \bar e^a_\mu-c_2 \bar f^a_\mu$.   The same linear combinations    appear as small perturbations in the $GL(2,C)$ invariant metric tensor ${\tt g}_{\mu\nu}$, since substituting (\ref{grvtnmds}) in (\ref{gltcmtrc}) gives
\be {\tt g}_{\mu\nu}= \eta_{\mu\nu} + \rho_{a\mu}\delta^a_\nu + \rho_{a\nu}\delta^a_\mu\; \ee   Thus, as in standard gravity theories, the metric tensor contains all the graviton modes.  The massive modes of the theory are present in ${\tt B}_{\mu\nu}$   and the higher order invariants (\ref{honvrnts}).

Finally, other invariants can be constructed from the  curvature and torsion, which for the $GL(2,C)$ gauge theory are
   \beqa   F_{\mu\nu}&=& \partial_{[\mu} A_{\nu]} + i[A_\mu, A_\nu]  \cr & &\cr
    T_{\mu\nu}&=& \partial_{[\mu} E_{\nu]} +i[ A_{[\mu}, E_{\nu ]}] \;, \eeqa respectively. 
 The former contains the  Lorentz curvature (\ref{lrnzcrvtr}) and two $U(1)$ curvatures
 \be F_{\mu\nu}=\frac 12  R^{ab}_{\mu\nu}\sigma_{ab} + \partial_{[\mu} a_{\nu]} \BI +i\partial_{[\mu} b_{\nu]}  \gamma_5\;\label{gltwocrvtr}\ee 
The latter can be decomposed according to 
 \be T_{\mu\nu} =  t^a_{\mu\nu}\gamma_a +u^a_{\mu\nu}\gamma_5\gamma_a
\;,\ee with torsion tensors $t^{a}_{\mu\nu}$ and $u^{a}_{\mu\nu}$ defined by 
\beqa
 t^a_{\mu\nu}&=&\partial_{[\mu} e^{a}_{\nu]}  + \omega^{a}_{\;\;b[\mu } e_{\nu] }^b +2  f^a_{[\mu} b_{\nu ]}
\cr & &\cr u^a_{\mu\nu}&=&\partial_{[\mu} f^{a}_{\nu ]}  + \omega^{a}_{\;\;b[ \mu} f_{\nu] }^b +2  e^a_{[ \mu} b_{\nu ]}\label{trsntu}\;,
\eeqa
 thus generalizing the Lorentz torsion (\ref{sltctrsn}). 
 Now the Bianchi identities are 
 \beqa \partial_{[\mu }F_{\nu\rho ]} + i [A_{[\mu}, F_{\nu\rho ]}]&=& 0 \cr & &\cr
 \partial_{[\mu }T_{\nu\rho ]} + i [A_{[\mu}, T_{\nu\rho ]}] + i [E_{[\mu}, F_{\nu\rho ]}] &=& 0\eeqa
 $GL(2,C)$ invariant field actions were constructed from the  curvature (\ref{gltwocrvtr}) and vierbeins (\ref{gl2cvrbns}) in \cite{Chamseddine:2003we}.  Here, however, we shall not be concerned with the dynamics of the fields, and rather treat then as external in the point particle action.

 \subsection{A simple particle action}
 
The action for a point particle should  possess the necessary symmetries, which here include invariance under $GL(2,C)$ gauge transformations, general coordinate transformations and reparametrizations of the evolution parameter.  It should also reduce to the usual coupling to gravity in the absence of the additional fields of  $GL(2,C)$ gauge theory, i.e., $a_\mu$, $b_\mu$ and $f^a_\mu$.   
 For a point particle with mass $m\ne 0$, an obvious choice  is
\be  {\cal S}_0 = \int d\tau \;{\cal L}_0\;, \qquad\quad  {\cal L}_0 = m \sqrt{-{\tt g}_{\mu\nu}(z)\dot z^\mu\dot z^\nu}\;,\label{mstobvsS}\ee
 where  $\dot z^\mu=\frac{d z^\mu}{d\tau}$, $z^\mu(\tau)$ being the particle's space-time coordinates and $\tau$  parametrizes its world line.  It possesses all of the required symmetries, and  reduces to the standard  action for a massive particle in the limit $f^a_{\mu}\rightarrow 0$.
  The Euler-Lagrange equations of motion
  correspond to equations of parallel transport for the vector  ${\cal L}_0 ^{-1} \dot z^\mu$ 
 \be \frac D{D\tau} \Bigl({\cal L}_0 ^{-1} {\dot z^\lambda} \Bigr)\;\equiv\; \frac d{d\tau} \Bigl({\cal L}_0 ^{-1}  {\dot z^\lambda}\Bigr)\; +\;{\tt \Gamma}^\lambda_{\mu\nu} \; \Bigl({\cal L}_0 ^{-1} {\dot z^\mu}\Bigr) {\dot z^\nu} \;=\;0\;, \ee
where $ {\tt \Gamma}^\lambda_{\mu\nu}$ are the Christoffel symbols  constructed from the metric tensor  ${\tt g}_{\mu\nu}$, provided that ${\tt g}_{\mu\nu}$ is nonsingular.   As usual, we can  perform a reparametrization such that the transformed ${\cal L}_0$ is a constant, thereby recovering the geodesic equations.
  This corresponds to transforming  $\tau$ to   the proper time, i.e., 
$d\tau^2\rightarrow -{\tt g}_{\mu\nu}(z)   {dz^\mu} {dz^\nu}$.
 
 Although it is reassuring that we recover geodesic motion, the $GL(2,C)$ gauge theory contains more degrees of freedom than is found in standard gravity theory, and so more particle interactions are possible. In addition to spin, the particle can have two $U(1)$ charges, say $q$ and $\tilde q$, associated with the two $U(1)$ gauge fields. The interaction terms
 \be \int d\tau  \;\Bigl(q a_\mu (z)+\tilde q b_\mu (z)\Bigr)\;\dot z^\mu \label{uonentrctn}\ee  can then be considered. Moreover, in addition to ${{\tt g}_{\mu\nu}(z)\dot z^\mu{\dot  z^\nu}}$, the total action can  involve  the higher order $GL(2,C)$ invariant ${{\tt h}_{\mu\nu\rho\sigma}(z)\dot z^\mu{\dot  z^\nu}{\dot z^\rho}{\dot z^\sigma}}$.  In what follows we  give a systematic approach to writing down particle dynamics in this theory, and show that the general action  contains such  higher order invariants terms, as well as interaction terms (\ref{uonentrctn}).
 
 \section{ Particle Classification}
 
 \setcounter{equation}{0}
 
 \subsection{Orbits in Momentum Space}
 
 Particles are standardly classified by the orbits which are traced out  in four-dimensional momentum space under the action of the Lorentz group.  If $p^a$ denotes the particle  momenta, then the  action is generated by the variations  \be \delta_\lambda (p^a\gamma_a) = i[p^a\gamma_a,\lambda]\;,\label{vrtnn4mspc}\ee where $\lambda=\frac 12 \lambda^{ab} \sigma_{ab}$. Six distinct orbits can be identified, only two of which are physically relevant and they correspond to  positive energy massive and massless particles. (See for example, \cite{Balachandran:1986ab},\cite{Weinberg:1995mt}.)
We now replace $p^a \gamma_a $ by some matrices $P$, the Lorentz group by $GL(2,C)$, and the   variations (\ref{vrtnn4mspc}) by 
 \be \delta_\Lambda  P = i[ P,\Lambda] \;,\label{adjntactnonP} \ee where $\Lambda$ was defined in (\ref{ALmbd}).  For closure we need that $P$ is a linear combination of both $\gamma_a$ and $\gamma_5\gamma_a$ matrices.  Thus momentum space must be enlarged to an eight-dimensional real vector space  ${\mathbb{R}}^8$ spanned by real vectors, say $p^a$ and $\tilde p^a$.
 (Alternatively, we can introduce the complex momentum  vector $p^a+i \tilde p^a$.) Upon writing 
\be P = p^a \gamma_a + \tilde p^a \gamma_5\gamma_a \;,\ee it follows that $p^a$ and $\tilde p^a$ transform under $GL(2,C)$ as the vierbeins $e^a_\mu$ and $f^a_\mu$ in (\ref{emuavrtn}) and (\ref{cmpsgtrns}), i.e.,
 \beqa  \delta_\Lambda p^a &=&  p^b \lambda_b^{\;\;a}         +2 \tilde p^a_\mu \beta\cr & &\cr
  \delta_\Lambda \tilde p^a &=&  \tilde p^b \lambda_b^{\;\;a}          +2 p^a \beta\;
\label{vrtnptldp}\eeqa  
 
 Many more distinct orbits  are possible upon enlarging  the momentum space to ${\mathbb{R}}^8$.  These orbits are generated  by the adjoint action (\ref{adjntactnonP}), and can be classified by their $GL(2,C)$ invariants.
  For this, it is convenient to re-express  $p^a$ and $\tilde p^a$  in terms of the following $2\times 2$  hermitean matrices ${\cal P}$ and $\bar {\cal P}$:
 \beqa {\cal P}&= &(p^0 +\tilde p^0)\BI + (p^i +\tilde p^i)\sigma_i \cr & &\cr
  \bar {\cal P}&= &(p^0 -\tilde p^0)\BI - (p^i -\tilde p^i)\sigma_i \;,\label{PbPshrmmtrx}  \eeqa  $\sigma_i$ being the Pauli matrices.  ${\cal P}$ and $\bar{\cal P}$ transform under  $GL(2,C)$ according to 
 \beqa {\cal  P}\rightarrow {\cal P}' &=& M{\cal P}M^\dagger \cr & &\cr \bar {\cal P}\rightarrow \bar {\cal P}'& =& {M^\dagger}^{-1}\bar {\cal  P} M^{-1} \;,\label{trnsfPbP}\eeqa where $M$ is a  $GL(2,C)$ matrix written in  the defining representation. This agrees with (\ref{vrtnptldp}) for infinitesimal transformations. The space of all ${\cal P}'$ and $\bar {\cal  P}'$ generated from ${\cal P}$ and $\bar {\cal  P}$  in (\ref{trnsfPbP}) defines an orbit. 
 Then 
 \beqa {\cal C}^{(2)}\;=\;  p_a p^a-\tilde p_a \tilde p^a& =&-\frac 12\;{\rm tr} {\cal P}\bar {\cal  P} \label{psquared}\\ & &\cr 
 {\cal C}^{(4)}\;=\;( p_a p^a+\tilde p_a \tilde p^a)^2 -4(p_a \tilde p^a)^2&=&  \det{\cal P}\det \bar {\cal  P}  \label{psquatic}
 \eeqa
 are quadratic and quartic invariants, respectively, and serve to label the orbit.  The former defines the invariant norm of the momenta and reduces to minus the mass-squared when $\tilde p^a\rightarrow 0$.  The latter can  be re-expressed as $ {\cal C}^{(4)}=\frac 12\;({\rm tr} {\cal P}\bar {\cal P})^2-\frac 12\;{\rm tr}({\cal P}\bar {\cal  P})^2$. 
 Although  neither \beqa \det {\cal P}&=&-(p^a+\tilde p^a)(p_a+\tilde p_a)\eeqa  nor\beqa\det\bar {\cal  P}&=&-(p^a-\tilde p^a)(p_a-\tilde p_a)\eeqa are invariant under $GL(2,C)$ transformations, their signs are; i.e.,  $\det {\cal P}$ and $\det \bar{\cal P}$ are either positive, negative or zero for all points on any orbit.  Thus they  can also be used to label the orbits.  Of course, for ${\cal C}^{(4)}\ne 0$  they are not independent.  There are therefore at least three invariants which can be used to classify the orbits.  (More will be obtained in section 7.2 upon including the spin.)
 
  For the action (\ref{mstobvsS}) considered previously, a reasonable choice for the `momenta' $p^a$ and $\tilde p^a$ is  
 \be p^a = \frac{m v^a}{\sqrt{ \tilde v^a \tilde v_a - v^a v_a}}\qquad\quad \tilde p^a = \frac{m \tilde v^a}{\sqrt{ \tilde v^a \tilde v_a - v^a v_a}} \;,\label{ptpmsvcs}\ee with `velocities' $v^a$ and $\tilde v^a$ given by
  \be
v^a=e^a_\mu(z)\dot z^\mu\qquad \quad \tilde v^a=f^a_\mu(z)\dot z^\mu \label{uvmstobvsS}\ee They  transform under the action of $GL(2,C)$ as  $p^a$ and $\tilde p^a$, respectively. Then $\tilde v^a \tilde v_a - v^a v_a$ is invariant and corresponds to $({\cal L}_0/m)^2$.
From the assignment (\ref{ptpmsvcs}), it follows that the canonical momenta  $\pi_\mu=\partial{\cal L}_0/\partial \dot z^\mu$ are equal to the linear combinations \be \pi_\mu=p_ae^a_\mu-\tilde p_af^a_\mu \label{canmom}\ee 
 For $p^a$ and $\tilde p^a$ defined this way,  ${\cal C}^{(2)}$ is minus the mass-squared, while the invariant  ${\cal C}^{(4)}$ is dynamically determined
 \be 
{\cal C}^{(2)}=-m^2 \qquad\qquad{\cal C}^{(4)}= m^4\;\Biggl(2-\frac {{\tt h}_{\mu\nu\rho\sigma}(z)\dot z^\mu\dot z^\nu\dot z^\rho\dot z^\sigma}{\Bigl(2 {\tt g}_{\eta\xi}(z)\dot z^\eta\dot z^\xi \Bigr)^2}\Biggr)\label{c4c4frsndL}\ee 
In the special  case where all  the $f^a_\mu$ vierbeins can be transformed away using a $GL(2,C)$ gauge transformation, then \be {{\tt h}_{\mu\nu\rho\sigma}(z)\dot z^\mu\dot z^\nu\dot z^\rho\dot z^\sigma}=\Bigl(2 {\tt g}_{\eta\xi}(z)\dot z^\eta\dot z^\xi \Bigr)^2\;,\label{agrewt1a}\ee and   ${\cal C}^{(4)}$ reduces to $m^4$.

 \subsection{General Classification}
 
  We now drop the definitions of $p^a$ and $\tilde p^a$ as given in  (\ref{ptpmsvcs}) and consider general orbits generated by (\ref{trnsfPbP}) in ${\mathbb{R}}^8$.  These orbits can be classified using the invariants ${\cal C}^{(2)}$, ${\cal C}^{(4)}$ and  $\det {\cal P}$ (and/or $ \det \bar {\cal  P}$).    As we shall see later, further  quantities are needed to classify orbits with ${\cal C}^{(2)} ={\cal C}^{(4)}= \det {\cal P}= \det \bar {\cal  P}=0$.
  
   General orbits are defined by the set of all $\{{\cal P}\}$ and $\{\bar {\cal P }\}$  with \be {\cal P}= NKN^\dagger \qquad\qquad \bar {\cal P }= {N^\dagger}^{-1}\bar K N^{-1} \;,\label{defPbarP}\ee where  $N$ denote   $GL(2,C)$ matrices written in the defining representation, while $K$ and $\bar K$ are constant $2\times 2$ hermitean matrices 
 \beqa K&= &(k^0 +\tilde k^0)\BI + (k^i +\tilde k^i)\sigma_i \cr & &\cr
  \bar K&= &(k^0 -\tilde k^0)\BI - (k^i -\tilde k^i)\sigma_i  \;,\eeqa
  which we can associate with a fiducial point $(k,\tilde k)$ on the orbit.
      ${\cal P}$ and $  \bar {\cal  P}$ are invariant under 
  \be  N\rightarrow N' =N e^{i\alpha} \;,\label{uonesbgrp}\ee corresponding to a $U(1)$ gauge symmetry.  More generally, there is a gauge symmetry associated with the right action on $N$ by the little group $G_{k,\tilde k}=\{n\}$ of both $K$
 and   $\bar K$: 
 \be N\rightarrow N' =Nn\;,\label{lggtrns}\ee where \be  nKn^\dagger =K\qquad{\rm and}\qquad {n^\dagger}^{-1}\bar K n^{-1} =\bar K\label{ltlgrpcnstr}\ee   As is usual, the little   groups  are isomorphic for all points on an orbit and can therefore be used to classify the orbits   $\{{\cal P}\}$ and $\{\bar {\cal P }\}$ in ${\mathbb{R}}^8$.

 Among the many possible orbits are those which have fiducial points $(k,0)$. If we restrict to  transformations by the $SL(2,C)$ subgroup of $GL(2,C)$, then provided $k\ne0$, the familiar  orbits for massive particles, massless particles and tachyons are swept out in the four-dimensional subspace of ${\mathbb{R}}^8$ spanned by $p$, while only a point at the origin results in the $\tilde p-$subspace.  For this reason, we shall identify  orbits resulting from  the full action of $GL(2,C)$ in ${\mathbb{R}}^8$ containing the fiducial point $(k,0)$ with massive particles, massless particles and tachyons, depending on the choice for $k$.

 \noindent  $ia)$ Massive particle: $k^a=m\delta^a_{0}$ and $\tilde k=0$, $m\ne0$.  For this case, the invariants satisfy ${\cal C}^{(2)}=-m^2 <0$, ${\cal C}^{(4)}=m^4$ and
 ${\rm sign} \det{\cal P}={\rm sign} \det \bar{\cal P}=+$. 
 (\ref{c4c4frsndL}) reduces to this case when (\ref{agrewt1a}) holds. Here  $K=\bar K=m\BI$, having identical little groups equal to $G_{(m,\vec 0),0}=U(2)$, and so ${\cal P}$ and $\bar{\cal P}$ both span $GL(2,C)/U(2)$.  As in the case  of orbits obtained under the action of just the Lorentz group, we can divide this case into two subcases with $m>0$ and $m<0$.  This is since there is no $GL(2,C)$ transformation (\ref{trnsfPbP}) that connects the two subcases.
 We examine  dynamics  in flat space-time in section seven, and  recover the usual massive particle system in this case.   This is despite the  presence of the two momentum vectors $p^a$ and $\tilde p^a$.

\noindent   $ib)$ Massless particle: $k=(\nu,0,0,\nu)$ and $\tilde k=0$, $\nu\ne 0$.   All of the invariants vanish in this case, ${\cal C}^{(2)}={\cal C}^{(4)}= \det{\cal P}= \det \bar{\cal P}=0$. Now $K=\nu(\BI+\sigma_3)$ and $\bar K=\nu(\BI-\sigma_3)$, which again have identical little groups, now  $G_{(\nu,0,0,\nu),0}=U(1)\times E(2)$.  The latter is generated by $\BI,\sigma_3,\sigma_1+i\sigma_2,\sigma_2-i\sigma_1$.  [Note from (\ref{ltlgrpcnstr}) that the little group acts differently on $K$ and $\bar K$.]  Now   ${\cal P}$ and $\bar{\cal P}$  both span $GL(2,C)/(U(1)\times E(2))$.  As with $ia)$, this case can be subdivided into  $\nu>0$ and $\nu<0$, since there is no $GL(2,C)$ transformation (\ref{trnsfPbP}) that connects the two subcases.
  
\noindent $ic)$ Tachyon: $k^a=\kappa\delta^a_{3}$ and $\tilde k=0$, $\kappa\ne 0$.   The invariants are $ {\cal C}^{(2)}=\kappa^2>0$, ${\cal C}^{(4)}=\kappa^4$ and
 ${\rm sign} \det {\cal P}={\rm sign} \det \bar {\cal P}=-$.  Here $K=-\bar K=\kappa\sigma_3$, and so there is again a common little group  $G_{(0,0,0,\kappa),0}=U(1)\times SO(2,1)$, generated by $\BI,\sigma_3,i\sigma_1,i\sigma_2$. Consequently, ${\cal P}$ and $\bar{\cal P}$  both span $GL(2,C)/(U(1)\times SO(2,1))$.
 
 Complementary to the previous cases, we can consider orbits having  fiducial points $(0,\tilde k)$. If we again restrict to  transformations by the $SL(2,C)$ subgroup of $GL(2,C)$, then provided $\tilde k\ne0$, the familiar  orbits for massive particles, massless particles and tachyons are  swept out in the four-dimensional subspace of ${\mathbb{R}}^8$ spanned by $\tilde p$, while only a point at the origin results in $ p-$subspace.  The various subcases result from different choices for $\tilde k$.
 
 \noindent  $iia)$  $k^a=0$, $\tilde k^a=\tilde m\delta^a_{0}$, $\tilde m\ne 0$.  This is the complement of $ia)$.  Now the invariants are ${\cal C}^{(2)}=\tilde m^2>0$, ${\cal C}^{(4)}=\tilde m^4$,
 ${\rm sign} \det{\cal P}={\rm sign} \det \bar{\cal P}=+$. Here the signs of ${\cal C}^{(2)}$ and ${\cal C}^{(4)}$ are the same as  $ic)$, but this case has opposite signs for $\det{\cal P}$  and $\det\bar{\cal  P}$ .      Therefore,  $ic)$ and $iia)$ define distinct orbits.
Now $K=-\bar K=\tilde m\BI$ and, as in case $ia)$, both have the little group $G_{0,(\tilde m,\vec 0)}=U(2)$, and  so  ${\cal P}$ and $\bar{\cal P}$ both span $GL(2,C)/U(2)$.   $\tilde m >0$ and $\tilde m <0$ correspond to disconnected orbits.  In section seven, by going to flat space-time we show the subcase $iia)$ to be unphysical. 

\noindent   $iib)$ $k^a=0$, $\tilde k^a=(\tilde \nu,0,0,\tilde \nu)$, $\tilde \nu\ne 0$.  This is the complement of $ib)$.  $\tilde \nu >0$ and $\tilde \nu <0$ correspond to disconnected orbits. As with  $ib)$, all invariants  vanish:  ${\cal C}^{(2)} ={\cal C}^{(4)}= \det{\cal P}= \det \bar{\cal P}=0$.  Furthermore,
 $K=\tilde \nu(\BI+\sigma_3)$ and $\bar  K=-\tilde \nu(\BI-\sigma_3)$ have identical little groups and they are  the  same as the little group for $ib)$, i.e., $G_{0,(\tilde \nu,0,0,\tilde \nu)}=U(1)\times E(2)$.  Despite having the same invariants and little group,   $ib)$ and   $iib)$  define distinct orbits.
 This is because there is no $GL(2,C)$ transformation (\ref{trnsfPbP}) from $K=\tilde \nu(\BI+\sigma_3),\bar  K=-\tilde \nu(\BI-\sigma_3)$ to $K=\tilde \nu(\BI+\sigma_3),\bar K=\tilde \nu(\BI-\sigma_3)$, as such a transformation would have to be in the little group of $K$, but not in the little group of $\bar K$.  Therefore, the  invariants and little groups are not sufficient to distinguish  all possible orbits.

\noindent $iic)$    $k^a=0,\;\tilde k^a=\tilde \kappa\delta^a_{3}$, $\tilde \kappa\ne0$.  The invariants are 
$ {\cal C}^{(2)}=-\tilde \kappa^2<0$,  ${\cal C}^{(4)}=\tilde \kappa^4$ and 
 ${\rm sign} \det{\cal P}={\rm sign} \det \bar{\cal P}=-$, so here the signs of ${\cal C}^{(2)}$ and ${\cal C}^{(4)}$ are the same as for the massive particle orbits $ia)$, but with opposite signs for $\det{\cal P}$  and $\det\bar{\cal  P}$ .  Now $K=\bar K=\tilde \kappa\sigma_3$, and the resulting little groups are $G_{0,(0,0,0,\tilde \kappa)}=U(1)\times SO(2,1)$, as was true for $ic)$.  So as in that case,  ${\cal P}$ and $\bar{\cal P}$  both span $GL(2,C)/(U(1)\times SO(2,1))$.
 
In all of the previous cases, $K$ and $\bar K$ had identical little groups and ${\cal P}$ and $\bar {\cal P}$ spanned identical orbits.  More generally, one can consider cases where $K$ and $\bar K$ each have  different little groups and therefore  ${\cal P}$ and $\bar{\cal P}$ span different orbits.  Two special example correspond to either ${\cal P}$ or $\bar{\cal P}$  vanishing, implying a trivial orbit for either ${\cal P}$ or $\bar{\cal P}$ .  These orbits have $k^a=\pm \tilde k^a$ and  ${\cal C}^{(2)} ={\cal C}^{(4)}= 0$. 
 
 \noindent $iii)$ $k^a= \tilde k^a$.  This implies  $K=2(k^0\BI + k^i \sigma_i)$ and $\bar K=0$, and hence $\det {\cal P}=-4 p_ap^a$ and $\det \bar{\cal P}=0$.  Three separate subcases can  then be  considered:  $iiia)$  $\det {\cal P} >0$ ,  $iiib)$  $\det {\cal P} =0$ and  $iiic)$  $\det {\cal P} <0$.  They have little groups $U(2)$, $U(1)\times E(2)$ and $U(1)\times SO(2,1)$, respectively.
 
 \noindent $iv)$ $k^a=- \tilde k^a$.  This implies $K=0$ and  $\bar  K=2(k^0\BI - k^i \sigma_i)$, and hence $\det {\cal P}=0$, along with  $\det \bar {\cal P}=-4 p_ap^a$.   Again  three separate subcases can  then be  considered:  $iiv)$  $\det \bar {\cal P} >0$ ,  $ivb)$  $\det\bar  {\cal P} =0$ and  $ivc)$  $\det\bar  {\cal P} <0$, having little groups $U(2)$, $U(1)\times E(2)$ and $U(1)\times SO(2,1)$, respectively.
 
 \noindent  The invariants and little groups for $iiib)$ and $ivb)$ agree  with cases $ib)$ and $iib)$.  However, all cases correspond to distinct orbits for ${\cal P}$ and $\bar{\cal P}$, as no $GL(2,C)$ transformations (\ref{trnsfPbP}) connect the different assignments for $K$ and $\bar K$.   In section seven, we shall show that all four subcases of orbits lead to the same dynamics in flat space-time, namely that of a massless particle. Actually, all $iii)$ and $iv)$ orbits are associated with massless particles.  More surprisingly,  $iic)$ also describes a massless particle, although it contains additional degrees of freedom.  
 
 We summarize the results for the various orbits in the table below.
 
 $$
 \left.\matrix{ 
 &   {\cal C}^{(2)} & {\cal C}^{(4)} &\det {\cal P}   & \det\bar {\cal P}  & G_{k,\tilde k} \cr
 & &&    &  &   \cr
ia) \qquad& -m^2 & m^4& +   & + & U(2)  \cr
ib) \qquad & 0 & 0 & 0   & 0 & U(1)\otimes E(2) \cr
ic) \qquad& \kappa^2 & \kappa^4& -   & - & U(1)\otimes SO(2,1)   \cr
iia) \qquad& \tilde m^2 & \tilde m^4& +   & + & U(2)  \cr
iib) \qquad & 0 & 0 & 0   & 0 & U(1)\otimes E(2) \cr
iic) \qquad& -\tilde \kappa^2 & \tilde \kappa^4& -   & - & U(1)\otimes SO(2,1)   \cr
iiia) \qquad& \tilde 0 & 0& +   & 0 & U(2)  \cr
iiib) \qquad & 0 & 0 & 0   & 0 & U(1)\otimes E(2) \cr
iiic) \qquad &  0  & 0 & -   & 0 & U(1)\otimes SO(2,1)   \cr
iva) \qquad& \tilde 0 & 0& 0   & + & U(2)  \cr
ivb) \qquad & 0 & 0 & 0   & 0 & U(1)\otimes E(2) \cr
ivc) \qquad &  0  & 0 & 0   & - & U(1)\otimes SO(2,1)  \cr
}\right.  $$

\bigskip
\centerline{\small{ Invariants and little groups for various orbits}}

\section{ Particle Dynamics}

\setcounter{equation}{0}

Here we write down a $GL(2,C)$ invariant particle action which applies for all nontrivial orbits.  The approach is along the lines of \cite{Balachandran:1979ha} which yielded  general $SL(2,C)$ invariant particle actions.  We also obtain the equations of motions and a general class of solutions.   
  
  \subsection{$GL(2,C)$ Invariant Lagrangians}
  
   The Lagrangian can be constructed from  real invariant bilinears for  $GL(2,C)$.  There are two such bilinears, each of which are associated with the norm (\ref{psquared}).
  To see this we  introduce another set of $2\times 2$ hermitean matrices, denoted by ${\cal V}$ and $
  \bar{\cal V}$, which are defined to transform, respectively, as ${\cal P}$ and $\bar {\cal  P}$ in (\ref{trnsfPbP}).  Then  both ${\rm tr}\;{\cal P}\bar{\cal V}$ and  ${\rm tr}\;{\cal V}\bar{\cal P}$ are invariant.  Moreover, they are proportional to (\ref{psquared}) when ${\cal V}= {\cal P}$ and $\bar {\cal V}= \bar {\cal P}$.
   Upon writing  \beqa {\cal V}&= &(v^0 +\tilde v^0)\BI + (v^i +\tilde v^i)\sigma_i \cr & &\cr
  \bar{\cal V}&= &(v^0 -\tilde v^0)\BI -(v^i -\tilde v^i)\sigma_i  
 \;,\label{VbarV}\eeqa along with  (\ref{PbPshrmmtrx}), then two independent invariant bilinears can be expressed as
    \beqa {\rm tr}\; ({\cal P}\bar {\cal V} +{\cal V}\bar{\cal P})&=& -4(p_a v^a-\tilde p_a \tilde v^a)\label{nvrntblnrone} \\
 & &\cr  {\rm tr}\;({\cal P}\bar{\cal V} -{\cal V}\bar{\cal P})&=&-4(\tilde p_a v^a- p_a \tilde v^a) \label{nvrntblnrtwo} \eeqa

   We shall use the definitions of  ${\cal P}$ and $   \bar {\cal P}$ as given in (\ref{defPbarP}) in writing down the general particle Lagrangian.  $ v^a$ and $\tilde v^a$ will denote the `velocities' defined in  (\ref{uvmstobvsS}).  The  matrices in (\ref{VbarV}) can then be expressed as 
 \be  {\cal V}={\cal E}_\mu(z)  \dot z^\mu\qquad\quad \bar {\cal V}=\bar {\cal E}_\mu(z) \dot z^\mu\;,\ee
 where ${\cal E}_\mu(x)$ and $\bar {\cal E}_\mu(x)$ are the  space-time dependent $2\times 2$ hermitean matrices
  \beqa{\cal  E}_\mu&= &(e_\mu^0 +f_\mu^0)\BI + (e_\mu^i +f_\mu^i)\sigma_i \cr & &\cr
  \bar {\cal E}_\mu &= &(e_\mu^0 -f_\mu^0)\BI -(e_\mu^i -f_\mu^i)\sigma_i  
 \;\label{EbarEmu}\eeqa 
 
   The particle Lagrangian ${\cal L}_K$ can be written down using the invariants (\ref{nvrntblnrone}) and (\ref{nvrntblnrtwo}).
 The particle degrees of freedom  in this case are $z^\mu(\tau)$,  $N(\tau)$ and  $N^\dagger(\tau)$.  
   A  general expression for the Lagrangian is $\rho\;{\rm tr}\;{\cal P}\bar{\cal V} +\bar\rho \;{\rm tr}\;{\cal V}\bar{\cal P}$, where $\rho$ and $\bar\rho$  are constants.  These constants can be  absorbed into the definitions of $K$ and $\bar K$,  respectively, and so without any loss generality we can define 
   \be {\cal L}_K = -\frac 14{\rm tr}\Bigl( NKN^\dagger\bar {\cal E}_\mu(z) +{N^\dagger}^{-1}\bar K N^{-1} {\cal E}_\mu(z)\Bigr) \dot z^\mu \label{prtcllgrn}\;\ee 
    In the case of the orbits $ia)$ for a massive particle and $ib)$ for a massless particle, the Lagrangian 
 (\ref{prtcllgrn}) reduces to 
 \beqa {\cal L}^{ia) }_K &=&-\frac m4\;{\rm tr}\Bigl(NN^\dagger\bar{\cal  E}_\mu(z) +(NN^\dagger)^{-1}{\cal   E}_\mu(z)\Bigr) \dot z^\mu \label{Lia}\\ & &\cr
  {\cal L}^{ib) }_K& =&-\frac \nu 4 \;{\rm tr}\Bigl(N(\BI+\sigma_3)N^\dagger\bar {\cal E}_\mu(z) +{N^\dagger}^{-1}(\BI-\sigma_3) N^{-1} {\cal E}_\mu(z)\Bigr) \dot z^\mu \;, \eeqa respectively.
 The corresponding particle action ${\cal S}_K =\int d\tau {\cal L}_K $ is invariant under reparametrizations $\tau\rightarrow\tau'= f(\tau)$ and transformations under the action of the little group (\ref{lggtrns}).  The  $GL(2,C)$ gauge symmetry appears upon treating the fields dynamically, with the associated gauge transformations: 
  \beqa N(\tau)&\rightarrow & N'(\tau)=M N(\tau)\cr & &\cr{\cal  E}_\mu(z)&\rightarrow &{\cal E}'_\mu(z) = M{\cal E}_\mu(z)M^\dagger\cr & &\cr \bar {\cal E}_\mu(z)&\rightarrow &\bar {\cal E}'_\mu(z) = {M^\dagger}^{-1}\bar {\cal E}_\mu(z) M^{-1}\;,\label{gl2cgspa}\eeqa where $M=M[z(\tau)]$ is  a  $GL(2,C)$ matrix.  The action is then also invariant under general coordinate transformations.

 \subsection{Equations of motion}
 
We next obtain the Euler-Lagrange equations which  follow from variations of $N$, $N^\dagger$ and $z^\mu$ in the Lagrangian (\ref{prtcllgrn}).  General variations of $N$  lead to 
 \be {\cal P}\bar{\cal  V}={\cal V}\bar {\cal P}  \label{PbVVbP}\;, \ee while variations of $N^\dagger$ lead to its hermitean conjugate.
   Upon expanding these equations of motion in `velocity' and `momentum' components one gets
 \beqa  v^{[a} p^{b]} - \tilde v^{[a} \tilde p^{b]} &=&0\label{vapb}\\& &\cr
 v^a\tilde p_a -\tilde v^a p_a &=&0 \label{cmntseom}\eeqa
 In general, these equations, along with (\ref{psquared})  and (\ref{psquatic}), may not uniquely determine $p^a$ and $\tilde p^a$ in terms of $v^a$ and $\tilde v^a$.  
 The Euler-Lagrange equations that  follow from  variations of $z^\mu$ in (\ref{prtcllgrn}) are 
 \be {\rm tr} \Bigl(\frac { d{\cal P}}{d\tau} \bar {\cal  E}_\mu  +\frac {d\bar {\cal P}}{d\tau} {\cal E}_\mu\Bigr) =  {\rm tr}(\bar{\cal  P}\partial_{[\mu}{\cal E}_{\nu]} + {\cal P}\partial_{[\mu}\bar {\cal E}_{\nu]}) \;\dot z^\nu\; \label{vrtnofz}\ee  These equations can be re-expressed in a covariant manner
 upon introducing the covariant derivatives
\beqa D_\tau {\cal P} &=&\frac { d{\cal P}}{d\tau}  + i ({\cal A}_\mu {\cal P} - {\cal P}{\cal A}_\mu^\dagger)\dot z^\mu\cr & &\cr
  D_\tau \bar{\cal P} &=&\frac { d\bar{\cal P}}{d\tau}  + i ({\cal A}_\mu^\dagger \bar{\cal P} - \bar{\cal P}{\cal A}_\mu)\dot z^\mu\;,\eeqa
  where  ${\cal A}_\mu$ is the  $GL(2,C)$ connection, now expressed in the defining representation.  It gauge transforms as 
\be {\cal A}_\mu\rightarrow {\cal A}'_\mu
 =M{\cal A}_\mu M^{-1} +i \partial_\mu M M^{-1}\;, \label{Adefrep} \ee 
 where $M=M(x)$ is  a  $GL(2,C)$ matrix.
 The infinitesimal version of (\ref{Adefrep}) was given in (\ref{gl2cgtrns}).
 In terms of component gauge potentials $\omega_\mu^{ab}$, $ b_\mu$ and $a_\mu $, ${\cal A}_\mu$ is given by
 \be  {\cal  A}_\mu = \frac 14 (\epsilon_{ijk}\omega_\mu^{ij}+  2i \omega_\mu^{0k})\sigma_k+ \frac 12(a_\mu+i b_\mu )\BI \;, \label{gltwoccntn}\ee
 Then  (\ref{vrtnofz}) can be re-written as
 \be {\rm tr} \Bigl( D_\tau {\cal P} \;\bar {\cal  E}_\mu  +D_\tau \bar {\cal P}\; {\cal E}_\mu \Bigr) =  {\rm tr}\Bigl(\bar{\cal  P}{\cal T}_{\mu\nu} + {\cal P} \bar {\cal T}_{\mu\nu}\Bigr) \;\dot z^\nu\;,\label{eomfrzmunspn}\ee  where we used equations of motion (\ref{PbVVbP}). ${\cal T}_{\mu\nu}$ and $ \bar {\cal T}_{\mu\nu}$ denote the $GL(2,C)$ generalization of the torsion, here written as $2\times 2$ hermitean matrices:
 \beqa {\cal T}_{\mu\nu}&=&\partial_{[\mu}{\cal E}_{\nu]} + i{\cal E}_{[\mu}{\cal A}^\dagger_{\nu]}-i {\cal A}_{[\nu}{\cal E}_{\mu ]}  \cr & &\cr
 \bar {\cal T}_{\mu\nu}&=&\partial_{[\mu}\bar{\cal E}_{\nu]} + i\bar{\cal E}_{[\mu}{\cal A}_{\nu ]}-i {\cal A}^\dagger_{[\nu }\bar{\cal E}_{\mu ]}\label{gl2ctrsn2by2}
 \eeqa
 They transform as  ${\cal E}_\mu(x)$ and $\bar {\cal E}_\mu(x)$, respectively, and hence the left and right hand sides of (\ref{eomfrzmunspn}) are invariant under $GL(2,C)$ gauge transformations.   ${\cal T}_{\mu\nu}$ and $ \bar {\cal T}_{\mu\nu}$  can also be expressed in terms of  the component torsion fields $t^{a}_{\mu\nu}$ and $u^{a}_{\mu\nu}$ which were defined in (\ref{trsntu}),  
 \beqa {\cal T}_{\mu\nu}&=&(t_{\mu\nu}^0 +u_{\mu\nu}^0)\BI + (t_{\mu\nu}^i +u_{\mu\nu}^i)\sigma_i 
    \cr & &\cr
 \bar {\cal T}_{\mu\nu}&=&(t_{\mu\nu}^0 -u_{\mu\nu}^0)\BI -(t_{\mu\nu}^i -u_{\mu\nu}^i)\sigma_i\;
 \eeqa
 The right hand side of (\ref{eomfrzmunspn}) vanishes for the case of zero torsion.
 In order for the covariant derivatives of ${\cal P}$ and $\bar {\cal P}$ to then vanish we would further need  the vierbein matrices $e^a_\mu$ and $f^a_\mu$ to be nonsingular and $e^a_\mu f^\mu_b =f^a_\mu e^\mu_b =0$, where $e^\mu_b$ and $ f^\mu_b $ are the inverses of $e^a_\mu$ and $f^a_\mu$, respectively.
 
 \subsection{Solutions}

 Here we first obtain a general class of solutions to equations of motion (\ref{PbVVbP}) which are valid when $ {\cal V}$  and $ \bar{\cal V}$ are nonsingular matrices.   We can apply the results to the various orbits discussed in section four.  For the choice $ia)$ associated with massive particles,  we obtain an effective Lagrangian, containing corrections to the naive Lagrangian  (\ref{mstobvsS}), and thus yielding corrections to geodesic motion. We  also find deviations from null curves for orbits $ib)$ associated with massless particles.  Finally, we examine the case of singular matrices $ {\cal V}$  and $ \bar{\cal V}$.  We are unable to find any physically meaningful solutions in that case. 
 
  \subsubsection{ $ \det {\cal V}\ne 0\;,\quad \det \bar{\cal V}\ne 0$}

  We first note that $ \bar {\cal V}^{-1} $ and ${\cal V}^{-1}$  transform under the action of $GL(2,C)$ as ${\cal P}$ and $ \bar{\cal P}$, respectively.
 So here since both  $ {\cal V}$  and $ \bar{\cal V}$ are nonsingular matrices, we may write  down the following solutions to (\ref{PbVVbP}):
 \beqa {\cal P}& =& \varsigma {\cal V} +\varpi \bar {\cal V}^{-1} \cr & &\cr \bar{\cal P} &=& \varsigma \bar {\cal V} +\varpi  {\cal V}^{-1} \;,\label{slnVVnv} \eeqa where $\varsigma$ and $\varpi$ are real and invariant under $GL(2,C)$ transformations.   For the special case where $\varpi= 0$ these solutions say that the `momenta' $p_a$ and $\tilde p_a$  are proportional to the `velocities' $v_a$ and $\tilde v_a$, as in (\ref{ptpmsvcs}).   More generally, $\varsigma$ and $\varpi$  are constrained by (\ref{psquared}) and (\ref{psquatic}).  Substituting (\ref{slnVVnv}) into these constraints gives rather involved conditions on $\varsigma$ and $\varpi$ 
 \beqa -2 {\cal C}^{(2)} &=& \biggl(\varsigma^2 +\frac{\varpi^2}{\det{(\cal V} {\bar {\cal V})}}\biggr){\rm tr} ({\cal V} \bar {\cal V}) + 4 \varsigma\varpi \cr & &\cr  {\cal C}^{(4)}  &=& \varsigma^4 {\det{(\cal V} {\bar {\cal V})}}
  +\frac {\varpi^4}{\det{(\cal V} {\bar {\cal V}})} + \varsigma^2\varpi^2 \biggl( \frac{ ({\rm tr} {\cal V} \bar {\cal V})^2}{\det{(\cal V} {\bar {\cal V})}} -6 \biggr) 
   - 4  \varsigma\varpi {\cal C}^{(2)}\;,\label{eqsfrvars}
 \eeqa 
 where we used the identity $ 2\det{(\cal V} \bar {\cal  V)}=  ({\rm tr} {\cal V}\bar {\cal V})^2-{\rm tr}({\cal V}\bar {\cal  V})^2$.  Solutions for $\varsigma$ and $\varpi$ can then in principle be expressed as functions of the invariants  ${\cal C}^{(2)}$, ${\cal C}^{(4)}$, 
\be {\rm tr} ({\cal V} \bar {\cal V}) = -2{\tt g}_{\mu\nu}(z)   {\dot z^\mu} {\dot z^\nu}\ee  and   
\be \det{(\cal V} {\bar {\cal V})} = \Bigl( 2 { {\tt g}_{\mu\nu}(z){\tt g}_{\rho\sigma}(z)}-
\frac 14 {{\tt h}_{\mu\nu\rho\sigma}(z)\Bigr)\dot z^\mu\dot z^\nu\dot z^\rho\dot z^\sigma} \label{dtVdtbrV}
\ee 
The solutions for $\varsigma$ and $\varpi$ are highly nontrivial for arbitrary values of  ${\cal C}^{(2)}$ and  ${\cal C}^{(4)}$.  They simplify considerably upon specifying particular orbits.
As an example, we now consider the orbits $ia)$ associated with  massive particles.  The calculations  depend on the values  ${\cal C}^{(2)}=-m^2$ and ${\cal C}^{(4)}=m^4$, but not  on ${\rm sign}\det {\cal P}$ and ${\rm sign} \det\bar {\cal P}$.  Therefore the results also apply for the orbits $iic)$, which in section 7.1 will be shown to correspond to {\it massless} particles.
There are two real solutions  for $\varsigma$ and $\varpi$ in this case:   \beqa \varsigma =\pm\frac m{\sqrt{ \;{\rm tr} ({\cal V} \bar {\cal V})+2 \Bigr({\det{(\cal V} {\bar {\cal V})}}  \Bigl)^{1/2}   }} &\qquad & \varpi= \pm\frac {m\;  \Bigl({\det{(\cal V} {\bar {\cal V})}} \Bigr)^{1/2} }{\sqrt{ \; {\rm tr} ({\cal V} \bar {\cal V}) +2 \Bigl({\det{(\cal V} {\bar {\cal V})}} \Bigr)^{1/2} }}\;,\qquad\label{slnonevrp}\eeqa  which is valid for $\quad {\det{(\cal V} {\bar {\cal V})}}\ge 0\;$, $\quad{\rm tr} ({\cal V} \bar {\cal V})+2 \Bigl({\det{(\cal V}{\bar {\cal V})}} \Bigr)^{1/2}  >0\;,$ and
 \beqa \varsigma =\pm\frac m{\sqrt{ \;{\rm tr} ({\cal V} \bar {\cal V})-2 \Bigr({\det{(\cal V} {\bar {\cal V})}}  \Bigl)^{1/2}   }}  &\qquad & \varpi = \mp\frac {m\;  \Bigl({\det{(\cal V} {\bar {\cal V})}} \Bigr)^{1/2} }{\sqrt{ \; {\rm tr} ({\cal V} \bar {\cal V}) -2 \Bigl({\det{(\cal V} {\bar {\cal V})}} \Bigr)^{1/2} }}\;,\qquad\label{slntwovrp}\eeqa  which is valid for $\quad{\det({\cal V} {\bar {\cal V})}}\ge 0\;$, $\quad{\rm tr} ({\cal V} \bar {\cal V})-2 \Bigl({\det{(\cal V} {\bar {\cal V})}} \Bigr)^{1/2}  >0$.  
 We can eliminate one of the solutions by demanding that they are well defined in the limit  $f^a_\mu\rightarrow 0$, since we recover the standard gravity theory in this limit.  This coincides with the condition (\ref{agrewt1a}), and   $\; \;{\rm tr} ({\cal V} \bar {\cal V})+2 \Bigl({\det({\cal V} {\bar {\cal V})}} \Bigr)^{1/2}  $ vanishes as a result.  The solutions (\ref{slnonevrp}) are singular in this limit and we reject them for that reason.
 From (\ref{slnVVnv}) we thus have
 \beqa {\cal P}& =& \pm m\; \frac{{\cal V}\; - \; \Bigl({\det({\cal V} {\bar {\cal V})}} \Bigr)^{1/2} \;
  \bar {\cal V}^{-1}}{{\sqrt{ \;{\rm tr} ({\cal V} \bar {\cal V})\;-\;2 \Bigr({\det({\cal V} {\bar {\cal V})}}  \Bigl)^{1/2}   }}} \cr & &\cr \bar{\cal P} &=& \pm m\; \frac{\bar {\cal V} \;-\;  \Bigl({\det({\cal V} {\bar {\cal V})}} \Bigr)^{1/2}  \; {\cal V}^{-1} }{{\sqrt{ \;{\rm tr} ({\cal V} \bar {\cal V})\;-\;2 \Bigr({\det({\cal V} {\bar {\cal V})}}  \Bigl)^{1/2}   }}}\label{slnptpnsglr}\eeqa
  Substituting this solution  back into the Lagrangian (\ref{Lia}) gives
 \beqa  {\cal L}^{ia) }_K &\propto & \sqrt{ \;{\rm tr} ({\cal V} \bar {\cal V})\;-\;2 \Bigr({\det({\cal V} {\bar {\cal V})}}  \Bigl)^{1/2}   }\cr & & \cr &\propto & \sqrt{-{\tt g}_{\mu\nu}(z)\dot z^\mu\dot z^\nu -\sqrt{\Bigl( 2 { {\tt g}_{\mu\nu}(z){\tt g}_{\rho\sigma}(z)}-
\frac 14 {{\tt h}_{\mu\nu\rho\sigma}(z)\Bigr)\dot z^\mu\dot z^\nu\dot z^\rho\dot z^\sigma} }} \label{efctvlgrngn} \eeqa 
It reduces to the naive Lagrangian (\ref{mstobvsS}) when $f^a_\mu\rightarrow 0$, as then
the condition (\ref{agrewt1a}) applies.  Geodesic motion is then recovered in this limit.  More generally, however, (\ref{efctvlgrngn}) will give corrections to geodesic motion.

There are no solutions to (\ref{eqsfrvars}) when ${\cal C}^{(2)}={\cal C}^{(4)}=0 $ for arbitrary  ${\rm tr} ({\cal V} \bar {\cal V})$ and $ \det{(\cal V} {\bar {\cal V})}$.  This case corresponds to the orbits $ib)$ for massless particles, as well as for orbits  $iib)$.     On the other hand, there can be consistent solutions when
 \be{\rm tr} ({\cal V} \bar {\cal V})\pm 2 \Bigl({\det{(\cal V}{\bar {\cal V})}} \Bigr)^{1/2}=0\ee 
 This condition coincides with null curves ${\tt g}_{\mu\nu}(z)\dot z^\mu\dot z^\nu\rightarrow 0$ in the limit that $f^a_\mu\rightarrow 0$, but can yield corrections to null curves when  $f^a_\mu\ne 0$.
 $\varsigma $  and $\varpi$ are not completely determined in this case, but are instead constrained by  $\;\varsigma +2\varpi/{\rm tr} ({\cal V} \bar {\cal V})=0\;,$ and so
 \be {\cal P}=\varsigma\Bigl(   {\cal V} -\frac 12 {\rm tr} ({\cal V} \bar {\cal V}) \;\bar {\cal V}^{-1}\Bigr) \quad\qquad \bar{\cal P}=\varsigma\Bigl(  \bar {\cal V}  -\frac 12 {\rm tr} ({\cal V} \bar {\cal V}) \;{\cal V}^{-1}\Bigr) \label{zromsssln} \ee 
 Substituting the solution  back into the Lagrangian (\ref{Lia}) this time gives zero.  
  The solutions (\ref{zromsssln}) do not apply for orbits  $iii)$ and  $v)$ where either $  {\cal P} $or $\bar{\cal P}$ vanish. For  orbits  $iii)$ we would need that 
 $ \bar {\cal V}  =\frac 12 {\rm tr} ({\cal V} \bar {\cal V}) \;{\cal V}^{-1}\;,$ 
 while for $iv)$ we need  ${\cal V} =\frac 12 {\rm tr} ({\cal V} \bar {\cal V}) \;\bar {\cal V}^{-1}$.  However, these are equivalent conditions when both $  {\cal V} $ and $\bar{\cal V}$ are nonsingular and hence (\ref{zromsssln}) reduce to trivial solutions. 
 
 There are alternative solutions which are only valid for orbits  $iii)$ and $iv)$.   We can set ${\cal P} = \varsigma {\cal V} +\varpi \bar {\cal V}^{-1} $ and $ \bar{\cal P}=0.$  Solutions of (\ref{PbVVbP}) then require that $\varsigma {\cal V} \bar {\cal V}+\varpi \BI=0\;.$  We get the same condition upon demanding that ${\cal P} =0$ and $\bar {\cal P}= \varsigma \bar {\cal V} +\varpi  {\cal V}^{-1} $  satisfies (\ref{PbVVbP}).   Thus, \be {\cal P}=\varsigma\Bigl(   {\cal V} -\frac 12 {\rm tr} ({\cal V} \bar {\cal V}) \;\bar {\cal V}^{-1}\Bigr) \quad\qquad \bar{\cal P}=0 \label{slnfriiirbt} \ee 
 and \be {\cal P}=0\qquad\quad \bar{\cal P}=\varsigma\Bigl(  \bar {\cal V}  -\frac 12 {\rm tr} ({\cal V} \bar {\cal V}) \;{\cal V}^{-1}\Bigr) \label{slnfrivrbt}\ee are solutions provided that 
 \be {\cal V} \bar {\cal V} =\frac 12 {\rm tr} ({\cal V} \bar {\cal V})\;\BI \label{532}\ee  
 The former applies for orbits $iii)$ and the latter for orbits $iv)$.

 \subsubsection{ $\det {\cal V}= 0\quad {\rm or}\quad\det \bar{\cal V}= 0$}
 
This case only applies when $ {\cal C}^{(4)} =0$.   Here we can in principle allow for $ {\cal C}^{(2)} \ne 0$, although we did not consider these kinds of orbits in sec. 4.2.
 With the exception of (\ref{slnfriiirbt}) and (\ref{slnfrivrbt}),  the  solutions obtained above are invalid for singular $ {\cal V}$ or $ \bar{\cal V}$. \footnote{ (\ref{slnfriiirbt}) is still  valid when   $ {\cal V}$ is singular and  (\ref{slnfrivrbt}) is valid when $ \bar{\cal V}$ is singular.}
The solutions (\ref{slnptpnsglr}) are ill-defined in the limit where either  $\det {\cal V}$ or $\det \bar{\cal V}$ (or both) vanish.  It follows that  (\ref{dtVdtbrV}) must also vanish in this limit.  Notice that this condition is different from (\ref{agrewt1a}), and unlike (\ref{agrewt1a}) it is not satisfied when $f^a_\mu\rightarrow 0$. 

If only ${\cal V}$ is singular we can write 
\be {\cal P} = \varsigma {\cal V} +\varpi \bar {\cal V}^{-1} \quad\qquad \bar{\cal P} = \bar\varsigma \bar {\cal V} \;,\ee while if only $\bar{\cal V}$ is singular we can have
\be {\cal P} = \varsigma {\cal V}  \qquad\quad \bar{\cal P} =\bar \varsigma \bar {\cal V} -\varpi  {\cal V}^{-1}\ee
They are solutions to the equation of motion (\ref{PbVVbP}) provided $(\bar \varsigma - \varsigma ) {\cal V}\bar {\cal V} = \varpi\BI$.  $\varsigma,\;\bar \varsigma$ and $\varpi$ are related by the quadratic invariant.  One gets ${\cal C}^{(2)}=\bar \varsigma^2\varpi /( \varsigma -\bar \varsigma )$ and ${\cal C}^{(2)}= \varsigma^2\varpi /( \varsigma -\bar \varsigma )$, respectively, for the two cases.

If, on the other hand, both  ${\cal V}$ and $\bar {\cal V}$ are singular, then the `momenta' ${\cal P}$ and $\bar{\cal P}$ are proportional to the `velocities' ${\cal V}$ and $\bar{\cal V}$,
\be  {\cal P} = \varsigma {\cal V} \quad\qquad\bar{\cal P} =\bar \varsigma \bar {\cal V} \;,\label{vrsgbrvrsg} \ee
 where here we need either             ${\cal V} \bar {\cal V}= 0$ or $\varsigma =\bar \varsigma $.
 In the former case ${\cal C}^{(2)}=0$, and in the latter $ {\cal C}^{(2)} =-\frac 12 { \varsigma^2}{\rm tr} ({\cal V} \bar {\cal V})$.

 Other special solutions to the equations of motion (\ref{cmntseom}) are
 \be\tilde p^a= \pm p^a\qquad \quad\tilde v^a= \pm v^a
 \label{tpeptvev}\;\ee   They correspond  to either $\bar{\cal  P}=\bar{\cal  V}=0$ or ${\cal P}={\cal V}=0$, and thus ${\cal C}^{(2)} = {\cal C}^{(4)} =0 $.  These solutions are relevant for the orbits $iii)$ and $iv)$.
However, (\ref{tpeptvev}) implies that $\Bigl(e^a_\mu(z)\mp f^a_\mu(z)\Bigr)\dot z^\mu=0$, which then also means  ${\tt g}_{\mu\nu}(z)\dot z^\mu=0$.  It follows that ${\tt g}_{\mu\nu}$ is a singular metric tensor and hence these solutions can only occur in a singular space-time.

\section{Wess-Zumino term}

\setcounter{equation}{0}

The dynamics discussed in the previous section only applies for particles with zero spin and zero charge.  It is known that the spin can be included with the addition of an $SL(2,C)$ invariant Wess-Zumino term.\cite{Balachandran:1979ha} It is first order in time derivatives of an $SL(2,C)$-valued matrix.   The term can easily be generalized to a $GL(2,C)$ invariant ${\cal L}_{WZ}$, which is first order in time derivatives of the $GL(2,C)$-valued matrices $N$ and $N^\dagger$.  The result is
\be {\cal L}_{WZ} = -\frac 14\;{\rm tr} \Bigl( W N^{-1} D_\tau N + W^\dagger (D_\tau N)^\dagger { N^\dagger}^{-1}\Bigr) \;,\label{wztrm}\ee   where $W$ is a constant $2\times 2$ complex  matrix. The covariant derivative in  (\ref{wztrm}) is given by  \be D_\tau N =\frac {d N}{d\tau} +i {\cal A}_\mu(z)N\dot z^\mu \;,\ee where ${\cal A}_\mu$ is again the $GL(2,C)$ connection (\ref{gltwoccntn}).  $W$ contains the six spin degrees of freedom of the particle in a fixed frame, along with two $U(1)$ charges. 
One can apply a similarity transformation on $W$ to go to an arbitrary reference frame:
\be \Sigma= NW N^{-1}=  (-i\epsilon_{ijk}s^{ij}+2 s^{0k})\sigma_k+ 2(\tilde q +i q)\BI \;,\label{SigmaNWNnvrs} \ee where $s^{ab}=-s^{ab}$ are the spin variables and $q$ and $\tilde q$ are the two $U(1)$ charges.
The former are
   dynamical quantities dependent on the traceless parts of $N$, while  $q$ and $\tilde q$ are constants, since Tr $ \Sigma\;=\;$Tr $W=   4(\tilde q +i q)$. The spin variables are unaffected by the action of the $U(1)$ subgroups of $GL(2,C)$.  So a general $GL(2,C)$ variation of $s^{ab}$ is just a $SL(2,C)$ variation
    \be  \delta_\Lambda s^{ab} =  { s}^{ac}\lambda_c^{\;\;b} -  {  s}^{bc}\lambda_c^{\;\;a} \ee
   Two Pauli-Lubanski-type vectors can be constructed for this theory
 \be w_a=\frac 12 \epsilon_{abcd} \;p^b s^{cd}\qquad\quad \tilde w_a=\frac 12 \epsilon_{abcd}\; \tilde p^b s^{cd} \;,\label{twoplvctrs}\ee
 which transform under  $GL(2,C)$ transformations as $p^a$ and $\tilde p^a$, respectively, in (\ref{vrtnptldp}).
 It follows that two additional invariants can then be constructed from 
 $w_a$ and $\tilde w_a$ which are
 analogous to (\ref{psquared}) and  (\ref{psquatic}):
 \beqa   {\cal C}^{(2)}_w&= &w_a w^a-\tilde w_a \tilde w^a\cr & &\cr 
{\cal C}^{(4)}_w &= &( w_a w^a+\tilde w_a \tilde w^a)^2 -4(w_a \tilde w^a)^2 \label{wwtldcsmr}
 \eeqa
 The former generalizes the usual  invariant for a relativistic spinning particle, while the second one is new.
 An additional invariant can be constructed from $w_a$, $\tilde w_a$,  $p^a$ and $\tilde p^a$,
  \be {\cal C}^{(2)}_{p,w}= \tilde p_a w^a- p_a \tilde w^a  = \epsilon_{abcd}\;\tilde p^a p^b s^{cd}\label{nvrntwthpnw}\ee
  Notice that the invariant $p_a w^a-\tilde p_a \tilde w^a$ is identically zero.
  (\ref{wwtldcsmr}) and (\ref{nvrntwthpnw}), along with
 (\ref{psquared}) and (\ref{psquatic}),  can be used to classify spinning particles in this theory. 
  More nontrivial invariants  using other combinations of  $w_a$, $\tilde w_a$,  $p^a$ and $\tilde p^a$ are may also be possible.

  The full Lagrangian for  spinning particles is  obtained by adding (\ref{wztrm}) to  the Lagrangian ${\cal L}_K$.  The corresponding action 
 \be {\cal S}= \int d\tau  \;{\cal L}\;,\quad\qquad {\cal L}={\cal L}_K + {\cal L}_{WZ} \;\label{ttlactn}\ee
 is gauge invariant and  reparametrization invariant.
 The gauge symmetries include transformations by the little group (\ref{lggtrns}), where now elements $\{n\}$ of the little group have to satisfy
 \be  nWn^{-1}=W\;, \ee which leaves (\ref{SigmaNWNnvrs}) invariant, in addition to the conditions (\ref{ltlgrpcnstr}).   If we treat the gauge fields dynamically, then the action (\ref{ttlactn}) is invariant under $GL(2,C)$ gauge transformations (\ref{gl2cgspa}) and  (\ref{Adefrep}).  In addition to the  $GL(2,C)$ gauge symmetries, the total action is invariant under independent $U(1)\times U(1)$ transformations, where the connections ${\cal A}_\mu$ transform as 
 \be {\cal A}_\mu\rightarrow {\cal A}'_\mu = {\cal A}_\mu + \partial_\mu\chi \;,\label{uoneuone}\ee
 for complex function $\chi$, while  $N$, ${\cal E}_\mu$ and $\bar{\cal E}_\mu$ are  unchanged.\footnote{Alternatively, we can  keep the connections ${\cal A}_\mu$  fixed, while  $N$, ${\cal E}_\mu$ and $\bar{\cal E}_\mu$ undergo the  transformations $$   N\rightarrow   e^\chi  N \qquad\quad {\cal  E}_\mu \rightarrow  e^{\chi + \chi^*}{\cal E}_\mu \qquad\quad \bar {\cal E}_\mu \rightarrow  e^{-\chi - \chi^*}\bar {\cal E}_\mu \nonumber $$}
 The Wess-Zumino Lagrangian picks up a $\tau$ derivative under such transformations.  
 
The Euler-Lagrange equations for the particle are again obtained  from variations of $N$, $N^\dagger$ and $z^\mu$. 
The equations of motion  which follow from  variations of $N$ in the total action (\ref{ttlactn}) are
 \be {\cal P}\bar{\cal  V}-{\cal V}\bar {\cal P}- D_\tau \Sigma = 0\;,\label{eomwthwatrm}\ee  generalizing (\ref{PbVVbP}), while variations of $N^\dagger $ gives its hermitean conjugate.  Here
 \be D_\tau \Sigma = \frac {d \Sigma}{d\tau} + i [{\cal A}_\mu(z),\Sigma]\;\dot z^\mu\ee  
 Upon expanding  in terms of components, (\ref{eomwthwatrm}) gives the the particle's spin precession
  \be  v^{[a} p^{b]} - \tilde v^{[a} \tilde p^{b]} +\dot s^{ab} +\omega^a_{\;\;c}s^{cb}-s^a_{\;\;c}\omega^{cb}=0\;,\label{wzcmntseom}\ee  while (\ref{cmntseom}) is unchanged.    The Euler-Lagrange equations following from variations of $z^\mu$ in the total action (\ref{ttlactn})  state that 
 \beqa & &{\rm tr} \Bigl(\frac { d{\cal P}}{d\tau} \bar {\cal  E}_\mu  +\frac {d\bar {\cal P}}{d\tau} {\cal E}_\mu+ i\frac {d\Sigma}{d\tau}{{\cal A}_\mu}- i\frac {d\Sigma^\dagger}{d\tau}{{\cal A}^\dagger_\mu}\Bigr) \cr & & \cr & &= \; {\rm tr}\Bigl(\bar{\cal  P}\partial_{[\mu}{\cal E}_{\nu]} + {\cal P}\partial_{[\mu}\bar {\cal E}_{\nu]}+ i \Sigma \partial_{[\mu}{\cal A}_{\nu]} -i \Sigma^\dagger \partial_{[\mu}{\cal A}^\dagger_{\nu]}\Bigr) \;\dot z^\nu\;,
 \eeqa or equivalently, in  the explicitly  gauge invariant form 
 \beqa 
 {\rm tr} \Bigl( D_\tau {\cal P} \;\bar {\cal  E}_\mu  +D_\tau \bar {\cal P}\; {\cal E}_\mu \Bigr)  &=&  {\rm tr}\Bigl(\bar{\cal  P}{\cal T}_{\mu\nu} + {\cal P} \bar {\cal T}_{\mu\nu}+i \Sigma {\cal F}_{\mu\nu}-i \Sigma^\dagger{\cal F}_{\mu\nu}^\dagger\Bigr) \;\dot z^\nu\;,\label{MPeqs}\eeqa generalizing (\ref{eomfrzmunspn}).  We have used the equations of motion  (\ref{eomwthwatrm}) in deriving (\ref{MPeqs}).   ${\cal T}_{\mu\nu}$ and $ \bar {\cal T}_{\mu\nu}$ again correspond to the $GL(2,C)$ torsion tensors (\ref{gl2ctrsn2by2}), while
  ${\cal F}_{\mu\nu}$ is the $GL(2,C)$ curvature, here expressed in the defining representation, i.e.,
  \be   {\cal  F}_{\mu\nu} = \frac 14 \Bigl(\epsilon_{ijk}R_{\mu\nu}^{ij}+2i R_{\mu\nu}^{0k}\Bigr)\sigma_k+ \frac 12\Bigl(\partial_{[\mu}a_{\nu]} +i\partial_{[\mu} b_{\nu]}\Bigr)\BI \;, \ee where  $R^{ab}_{\mu\nu}$ is the standard Lorentz curvature.  (\ref{MPeqs}) generalizes the Mathisson-Papapetrou equations \cite{Math},\cite{Papapetrou:1951pa}, by including interactions with the  two 
$U(1)$ gauge fields along with spin curvature and torsion.

\section{Flat space-time}

\setcounter{equation}{0}

  Flat space-time corresponds to the choice (\ref{fltLmu}) for the vierbein matrix $L_\mu$, or equivalently 
\be e^a_\mu =\cosh\chi\;\delta^a_\mu\qquad\quad
f^a_\mu =\sinh\chi\;\delta^a_\mu\;,\label{cnstendf}\ee for some real constant $\chi$.  
So in this case ${\cal V}$ and $
  \bar{\cal V}$ are given by \beqa {\cal V}&= &(\cosh\chi +\sinh\chi)\Bigl( \dot z^0\BI + \dot z^i\sigma_i \Bigr)\cr & &\cr
  \bar{\cal V}&= & (\cosh\chi -\sinh\chi)\Bigl( \dot z^0\BI- \dot z^i\sigma_i\Bigr)
 \;,\label{clVbeVfltspc}\eeqa and then $ {\cal V}
  \bar{\cal V} = - \dot z^a\dot z_a \;\BI $.   It follows that
 ${\rm tr}( {\cal V}
  \bar{\cal V}) =-2\dot z^a\dot z_a\;,\quad\det ({\cal V}
  \bar{\cal V}) =(\dot z^a\dot z_a)^2$ and furthermore that the condition (\ref{532}) is satisfied. 

By going to flat space-time we are breaking the $GL(2,C)$ gauge invariance of the Lagrangian.
Although this gauge invariance is broken, a number of symmetries survive.  They correspond to global Poincar\'e transformations, reparametrizations and the local  transformations  (\ref{lggtrns}).  Of course, the discrete symmetries, parity and time reversal, are present as well.  If we treat the two $U(1)$ potentials $a_\mu$ and $b_\mu$ dynamically, then the additional $U(1)\times U(1)$ gauge symmetry (\ref{uoneuone}) can also be included. 

Below we first consider the case of spinless and chargeless particles, and then remark on the inclusion of the Wess-Zumino term.

\subsection{Spinless and Chargeless particles}

In flat space-time the Lagrangian (\ref{prtcllgrn}) reduces to
\be {\cal L}_K =\pi_a\dot z^a\;,\qquad\quad \pi_a=\cosh\chi\;p_a -\sinh\chi\;\tilde p_a \label{LKinfltspc}\ee 
$\pi_a$ are the canonical momenta (\ref{canmom}), and from  the equations of motion (\ref{vrtnofz}), $\pi_a$ also serves as the conserved energy-momentum vector.  
Here the equations of motion  
 (\ref{vapb})  imply that   $z^{[a} \pi^{b]}$ are the conserved  angular momenta, and by Noether's theorem these two conservation laws are associated with the Poincar\'e symmetry.  We thus recover the standard dynamics for a free spinless particle.  On the other hand, this system, at first glance, contains additional degrees of freedom, as there are two momentum variables $p^a$ and $\tilde p^a$, or equivalently $\pi^a$ and 
 \be \tilde \pi _a=\cosh\chi\;\tilde p_a -\sinh\chi\; p_a \;\ee   The  dynamics of the latter is constrained by the   additional equation of motion  
 \be  \tilde \pi_a\dot z^a=0 \;, \label{uonecotm}\ee
 which follows  from (\ref{cmntseom}).
 $\pi^a$ and $\tilde \pi^a$ transform under $GL(2,C)$ transformations as $p^a$ and $\tilde p^a$, respectively, in (\ref{vrtnptldp}).
 The  quadratic and quartics invariants (\ref{psquared}) and (\ref{psquatic}) may be expressed directly in terms of $\pi_a$ and $\tilde\pi_a$, 
 \beqa {\cal C}^{(2)}&=&  \pi_a \pi^a-\tilde \pi_a \tilde \pi^a\cr & &\cr 
 {\cal C}^{(4)}&=&( \pi_a \pi^a+\tilde \pi_a \tilde \pi^a)^2 -4(\pi_a \tilde \pi^a)^2  \eeqa
 From the conservation of  angular momentum, it follows that $\pi_a\propto\dot z_a$, and then from (\ref{uonecotm}) that ${\cal C}^{(4)}\ge 0$ [assuming $\dot z\ne 0$].  Furthermore, 
 \beqa  \pi_a \pi^a&=&\frac 12\Bigl({\cal C}^{(2)}\pm\sqrt{{\cal C}^{(4)}}  \Bigr)\cr & &\cr  \tilde \pi_a\tilde  \pi^a&=&\frac 12\Bigl(- {\cal C}^{(2)}\pm\sqrt{{\cal C}^{(4)}} \Bigr)\cr & &\cr\pi_a\tilde  \pi^a&=& 0\;\label{cntrntsnpitpi}\eeqa
  
  We next examine these constraints for the various orbits discussed in section  four. 
 
  For orbits  $ib)$, $iib)$, $iii)$ and $iv)$, we have $ {\cal C}^{(2)}={\cal C}^{(4)}=0$  and so all scalar products in (\ref{cntrntsnpitpi}) vanish.  It follows that 
  $\tilde \pi^a=\lambda \pi^a$, and hence that $\tilde \pi^a$ are not  independent variables.  The independent degrees of freedom are those of a  massless particle, and this is valid for all four types of orbits.   The different orbits are distinguished by their values for $\lambda$.  These values can be determined from  the expressions for $p^a$ and $\tilde p^a$, 
  \be p_a=(\cosh\chi +\lambda \sinh\chi)\pi_a \qquad\quad \tilde p_a=(\sinh\chi +\lambda \cosh\chi)\pi_a\label{78}\ee
    The orbit $ib)$ is recovered for $\lambda=-\tanh\chi$ and $iib)$ is recovered for $\lambda=-1/\tanh\chi$.  
  $\lambda=1$  for  $iii)$ and $\lambda=-1$  for  $iv)$.   All of them correspond to the solutions (\ref{vrsgbrvrsg}) with  ${\cal V} \bar {\cal V}= 0$ with different values for $\varsigma$ and $\bar \varsigma $ in the four cases.

  For orbits  $ia)$, $ic)$, $iia)$ and $iic)$ where $ {\cal C}^{(2)}$ or ${\cal C}^{(4)}$ differ from zero, 
the choice of the sign in front of the  $\sqrt{{\cal C}^{(4)}}$ terms in (\ref{cntrntsnpitpi}) may be determined  from the signs of the determinants of ${\cal P}$ and $\bar{\cal P}$.  From (\ref{cntrntsnpitpi}) and (\ref{78}) one gets
\beqa \det {\cal P}&=& \mp(\cosh\chi+ \sinh\chi)^2\;\sqrt{{\cal C}^{(4)}}\cr& &\cr\det\bar{\cal P}&=&\mp (\cosh\chi- \sinh\chi)^2\;\sqrt{{\cal C}^{(4)}}\;\eeqa The signs of $\det{\cal P}$ and $\det\bar{\cal P}$  agree as is the case with all four classes of orbits.
For orbits $ia)$ and $iia)$ we must choose the lower sign and for orbits $ic)$ and $iic)$ we must choose the upper sign.  We now examine the flat space-time dynamics for  the four different orbits.

\noindent
 For the case $ia)$ of a massive particle where  ${\cal C}^{(2)}=-m^2$ and ${\cal C}^{(4)}=m^4$, the choice of the lower sign in (\ref{cntrntsnpitpi}) leads to the physically reasonable results, i.e.,  $ \pi_a \pi^a=-m^2$,  $ \pi_a\tilde  \pi^a=0$ and $\tilde \pi_a\tilde  \pi^a=0$.   The latter two equations mean that $\tilde \pi$ must vanish. [This is easily seen by going to  the particle rest frame   $\pi=( m,0,0,0)$.] Hence 
 \be  \pi^a =\pm \frac {m\dot z^a}{\sqrt{-\dot z^b\dot z_b}} \qquad\qquad\qquad \tilde \pi^a =0\;,\label{uslmsvfp}\ee  and so the independent degrees of freedom are those of a  massive particle.
The result also follows from the solution (\ref{slnVVnv}).  Upon substituting (\ref{clVbeVfltspc})  we get 
\beqa {\cal P}&= &\pm (\cosh\chi +\sinh\chi)\frac {m\dot z^a}{\sqrt{-\dot z^b\dot z_b}}\Bigl( \dot z^0\BI + \dot z^i\sigma_i \Bigr)\cr & &\cr
  \bar{\cal P}&= &\pm (\cosh\chi -\sinh\chi)\frac {m\dot z^a}{\sqrt{-\dot z^b\dot z_b}} \Bigl( \dot z^0\BI- \dot z^i\sigma_i\Bigr)\;,\eeqa from which follows
 \be p^a =\pm \cosh\chi\;\frac {m\dot z^a}{\sqrt{-\dot z^b\dot z_b}} \qquad\quad \tilde p^a =\pm \sinh\chi\;\frac {m\dot z^a}{\sqrt{-\dot z^b\dot z_b}}\;,\ee
 and hence (\ref{uslmsvfp}).

\noindent
 For case $iia)$ the invariants are ${\cal C}^{(2)}=\tilde m^2$ and ${\cal C}^{(4)}=\tilde m^4$.  So now $ \pi_a \pi^a=0$ and  $\tilde \pi_a\tilde  \pi^a=-\tilde m^2$.  The former implies that $\dot z^a$ is light-like or zero, while the latter means that $\tilde \pi^a$ is time-like.  But along with $ \tilde \pi_a \pi^a=0$,  this implies that $\pi^a=0$ and the $\dot z^a=0$.  [For this one can transform to the frame where  $\tilde \pi=(\tilde m,0,0,0)$.]   This therefore appears to be a pathological case.
 
\noindent
 For case $ic)$ one  has  $ {\cal C}^{(2)}=\kappa^2$ and  ${\cal C}^{(4)}=\kappa^4$. Upon choosing the upper sign in (\ref{cntrntsnpitpi}),  $ \pi_a \pi^a=\kappa^2$ and  $\tilde \pi_a\tilde  \pi^a=0$.  The result is  a tachyon, but here $\tilde \pi^a$ need not vanish.  
 
 \noindent
 For case  $iic)$, $ {\cal C}^{(2)}=-\tilde \kappa^2$ and  ${\cal C}^{(4)}=\tilde \kappa^4$.  Again choosing the upper sign in (\ref{cntrntsnpitpi}), one gets $ \pi_a \pi^a=0 $ and $\tilde \pi_a\tilde  \pi^a=\tilde \kappa^2$.  The former implies that the particle  velocity vector is light-like or zero, while the latter means that $\tilde \pi^a$ is space-like.  Here $\tilde \pi^a$ does not vanish, and also $\pi^a$ need not vanish.  The system therefore describes a massless particle.  This result is unexpected since the orbits here have  the same values for the invariants $ {\cal C}^{(2)}$ and   as $ {\cal C}^{(4)}$ as with the case of the massive particle $ia)$.  Unlike the massless particle orbits  $ib)$, $iib)$, $iii)$ and $iv)$, extra degrees of freedom are present for case $iiv)$, which are associated with the orthogonal space-like vector  $\tilde \pi^a$.

 \subsection{Inclusion of the Wess-Zumino term}

 The addition of the Wess-Zumino term (\ref{wztrm}),  with ${\cal A}_{\mu}=0$, to the total Lagrangian does does not affect the equations of motion $\dot\pi^a=0$ or (\ref{uonecotm}).  
 On the other hand, the addition of the Wess-Zumino term does lead to the inclusion of spin in the conserved angular momentum
\be j^{ab}=z^{[a} \pi^{b]}+s^{ab}\label{ttlanglmom} \ee      Infinitesimal Lorentz variations of $j^{ab}$ are as usual,
\be  \delta_\Lambda j^{ab} =  { j}^{ac}\lambda_c^{\;\;b} -  {  j}^{bc}\lambda_c^{\;\;a}\label{glbljab} \ee
Thus when ${\cal A}_{\mu}=0$, the $GL(2,C)$ Wess-Zumino term  (\ref{wztrm}) is equivalent to the $SL(2,C)$ Wess-Zumino term, and it only gives dynamics to the spin variables.\cite{Balachandran:1979ha} This is  evident because the Wess-Zumino term does not depend on the determinant of the $GL(2,C)$ matrix $N$ when ${\cal A}_{\mu}=0$.  $N$ can be   decomposed according to $N=\zeta \hat N$, with $\hat N\in SL(2,C)$ and the  terms in (\ref{wztrm}) (with ${\cal A}_{\mu}=0$) involving $\zeta$ are $\tau-$derivatives. 
If one assumes that the spin and orbital angular momentum  are separately conserved then the analysis of the motion for  orbits $i)-iv)$ is identical to what was found in section 7.1.

 Lastly, if we again consider flat space-time but now drop the restriction  that ${\cal A}_{\mu}=0$, the particle can feel the presence of Lorentz forces.  Upon  allowing for  the two $U(1)$ potentials; i.e.,  ${\cal A}_{\mu}=\frac 12 (a_\mu +i b_\mu)$, then the Wess-Zumino action will contain the minimal coupling terms  (\ref{uonentrctn}).  Although (\ref{uonecotm}) still holds, the momenta $\pi_a$ and angular momenta $j^{ab}$ are not in general conserved.  Rather, a Lorentz force equation results from the two gauge fields 
  \be \dot \pi_\mu=\Bigl(q\partial_{[\mu}a_{\nu]} + \tilde q\partial_{[\mu}b_{\nu]} \Bigr) \dot z^\nu \;\ee

 \section{Quantum theory}
 
 \setcounter{equation}{0}
 
Standard constraint Hamiltonian formalism can be applied to the Lagrangians of sections five and six in order to obtain the quantum theory.   The analysis proceeds in a similar fashion as that carried out in  \cite{Balachandran:1979ha} (second reference).  As  a multitude of constraints on the phase space  result in this case and the orbits have to studied separately, the procedure is quite lengthy.  Here we instead write down the quantum algebra which should result for all orbits, and sketch their  representations on momentum eigenstates.   The  algebra is  $16-$dimensional generalization of  the Poincar\'e algebra,  spanned by the two sets of momenta and the $GL(2,C)$ generators.    Unitary  representations of the algebra can be constructed along the lines of induced representations.
 
 For the quantum theory we  replace the two momentum vectors $p^a$ and $\tilde p^a$, respectively with the hermitian operators ${\bf p}^a$ and $ {\bf {\tilde p}}^a$, acting on  a Hilbert space ${\cal H}$. Additional observables are the $GL(2,C)$ generators  ${\bf  j}^{ab}=-{\bf  j}^{ba}$, ${\bf Y}$ and ${\bf  Z}$, where ${\bf j}^{ab}$ are the Lorentz generators and ${\bf Y}$ and ${\bf Z}$ are the $U(1)$ generators.  Since the generators are hermitean, we can construct the unitary operators 
 \be {\bf U}(\Lambda)=\exp\Bigl\{\frac i2 \lambda_{ab} {\bf j}^{ab} + i\alpha {\bf Y}+\beta {\bf Z} \Bigr\}\;, \ee
 for real parameters $\Lambda=( \lambda_{ab},\alpha,\beta)$.  The adjoint action   can then be utilized to induce  $GL(2,C)$ transformations on the space of observables $\{{\bf A}\}$,
 \be {\bf  A }\rightarrow {\bf  A}' ={\bf U}(\Lambda)^\dagger  {\bf A}\; {\bf U}(\Lambda)\ee
 For infinitesimal $\Lambda$, the transformations on  ${\bf  p}^a$, ${\bf {\tilde p}}^a$  and ${\bf  j}^{ab}$ are given in (\ref{vrtnptldp}) and (\ref{glbljab}).  Thus
 \beqa  {\bf U}(\Lambda)^\dagger  {\bf  j}^{ab}\; {\bf U}(\Lambda) &=&  {\bf j}^{ab}  +  {\bf j}^{ac}\lambda_c^{\;\;b} -  {\bf  j}^{bc}\lambda_c^{\;\;a} \cr & &\cr
  {\bf U}(\Lambda)^\dagger  {\bf p}^a\; {\bf U}(\Lambda) &=& {\bf p}^a  +  {\bf  p}^b \lambda_b^{\;\;a}             +2 {\bf {\tilde  p}}^a  \beta  \cr & &\cr
 {\bf  U}(\Lambda)^\dagger  {\bf {\tilde p}}^a\; {\bf U}(\Lambda) &=&  {\bf {\tilde p}}^a+ {\bf {\tilde p}}^b \lambda_b^{\;\;a}         +2 {\bf p}^a  \beta \;,
  \eeqa while ${\bf Y}$ and ${\bf Z}$ are invariant under $GL(2,C)$ transformations.  From this we then get the 
  quantum algebra for the observables ${\bf p}^a$, $ {\bf {\tilde p}}^a$,  ${\bf  j}^{ab}$, ${\bf Y}$ and ${\bf  Z}$.
    The nonvanishing commutators are
  \beqa i[ {\bf  j}^{ab},{\bf  j}^{cd}] &=& {\bf  j}^{ac}\eta^{bd}-{\bf  j}^{bc}\eta^{ad}-{\bf  j}^{ad}\eta^{bc}+{\bf  j}^{bd}\eta^{ac}\cr & &\cr
  i[ {\bf  p}^{a},{\bf  j}^{bc}] &=& {\bf  p}^{b}\eta^{ac}-{\bf  p}^{c}\eta^{ab}
  \cr & &\cr
  i[ {\bf {\tilde p}}^{a},{\bf  j}^{bc}] &=& {\bf {\tilde p}}^{b}\eta^{ac}-{\bf {\tilde p}}^{c}\eta^{ab}
  \cr & &\cr
  i[ {\bf  p}^{a},{\bf Z}] &=& 2{\bf {\tilde p}}^{a}
 \cr & &\cr
  i[ {\bf {\tilde p}}^{a},{\bf Z }] &=& 2{\bf  p}^{a}
  \eeqa
   The subgroup  generated by  ${\bf  j}^{ab}$ and  any linear combination of ${\bf p}^a$  and $ {\bf {\tilde p}}^a$ is  the Poincar\'e group.  [From (\ref{LKinfltspc}), the generator of translations in flat space-time is the linear combination $\cosh\chi\;{\bf p}_a -\sinh\chi\;{\bf \tilde p}_a$ .]
   ${\bf Y}$ is a central element, while ${\bf Z}$, ${\bf p}^a$ and ${\bf {\tilde p}}^a$ form a Euclidean algebra for fixed $a$.  The operator analogues of  
   ${\cal C}^{(2)}$ and ${\cal C}^{(4)}$ defined in (\ref{psquared})  and (\ref{psquatic}), along with ${\cal C}^{(2)}_w$, ${\cal C}^{(4)}_w$ and ${\cal C}^{(2)}_{p,w}$ in (\ref{wwtldcsmr}) and (\ref{nvrntwthpnw}), are Casimir operators whose values are fixed in any unitary irreducible representation.
   
Quantum states can be expressed in terms of eigenvectors $\Psi_{(p,\tilde p),\sigma,q}$ of    ${\bf p}^a$, ${\bf {\tilde p}}^a$ and  ${\bf Y}$.  (We cannot  include  ${\bf Z }$ in the set of commuting operators, and so the momentum eigenstates are labeled by a single charge.)
 \beqa {\bf p}^a \Psi_{(p,\tilde p),\sigma,q} &=&{ p}^a \Psi_{(p,\tilde p),\sigma,q}\cr & &\cr {\bf \tilde  p}^a \Psi_{(p,\tilde p),\sigma,q} &=&{ \tilde p}^a \Psi_{(p,\tilde p),\sigma,q}\cr & &\cr {\bf   Y} \Psi_{(p,\tilde p),\sigma,q} &=&{ q} \Psi_{(p,\tilde p),\sigma,q}  \eeqa
 $\sigma $ denote  degeneracy indices.   Following the usual procedure  for induced representations (see for example, \cite{Balachandran:1986ab},\cite{Weinberg:1995mt}),
 one can determine the spectrum for $\sigma$ by going to a fiducial point $(p,\tilde p)=(k,\tilde k)$ on an orbit, and then acting on the state $\Psi_{(k,\tilde k),\sigma,q}$ with elements of the little group 
 $G_{k,\tilde k}=\{n\}$ defined  in (\ref{lggtrns}).  Let $\{D^q_{\sigma' \sigma}\}$ be a unitary irreducible representation of $G_{k,\tilde k}$, which we define using
 \be {\bf U}(n)\Psi_{(k,\tilde k),\sigma,q}  = D^q_{\sigma' \sigma}(n)\Psi_{(k,\tilde k),\sigma',q}\;\ee
   One can then transform from the states at the fiducial point $(k,\tilde k)$ to  states at an arbitrary point  $(p,\tilde p)$ on the orbit using the analogue of a Wigner boost $L_{(p,\tilde p)}\in GL(2,C)$, where 
$ (p,\tilde p) =L_{(p,\tilde p)}\circ (k,\tilde k)$ and   $L_{(k,\tilde k)}$ is the identity map. Here $\circ$ denotes the action of $GL(2,C)$ on $(p,\tilde p)$ given by (\ref{trnsfPbP}). The states can be defined to transform as,
\be \Psi_{(p,\tilde p),\sigma,q} = {\cal N}_{(p,\tilde p)} {\bf U}(L_{(p,\tilde p)})\Psi_{(k,\tilde k),\sigma,q}\;,\ee  where  ${\cal N}_{(p,\tilde p)}$ is a normalization factor.   An arbitrary  transformation by $\Lambda\in GL(2,C)$ on this state is given by
\be {\bf U}(\Lambda )\Psi_{(p,\tilde p),\sigma,q}=\frac {{\cal N}_{(p,\tilde p)}}{{\cal N}_{\Lambda\circ(p,\tilde p)}}D^q_{\sigma' \sigma}(n_{\Lambda, (p,\tilde p)})\Psi_{\Lambda\circ(p,\tilde p),\sigma',q}  \;, \ee
where $n_{\Lambda, (p,\tilde p)}\in G_{k,\tilde k}$ is the analogue of a Wigner rotation,
\be  n_{\Lambda, (p,\tilde p)}= L_{\Lambda\circ(p,\tilde p)}^{-1} \circ \Lambda \circ L_{(p,\tilde p)}  \ee

\section{Discussion}
 
 \setcounter{equation}{0}

Using the quadratic invariants for $GL(2,C)$ we wrote down a general coupling of particles to an extended theory of gravity based on   $GL(2,C)$ gauge theory.\cite{Chamseddine:2003we}
Two momentum variables $p^a$ and $\tilde  p^a$ were needed to couple to the two sets of vierbein fields  $e^a_\mu$ and  $f^a_\mu$.
  We classified the orbits of these momenta  using the values of the three invariants ${\cal C}^{(2)}$, ${\cal C}^{(4)}$ and 
 ${\rm sign} \det{\cal P}$, 
 in addition to  their little groups $G_{k,\tilde k}$.  Various  orbits  were examined, which are summarized in the table at the end of section four. 
Further divisions of orbits can also be made based on the sign of the energy.  The orbits $ia)$  were identified with   massive particles, while $ic)$ represented  tachyons.  A degeneracy in  the classification was found when ${\cal C}^{(2)}={\cal C}^{(4)}=\det {\cal P}=0$ and $G_{k,\tilde k}=U(1)\otimes E(2)$, as  $ib)$, $iib)$, $iiib)$ and $ivb)$ represent  disconnected regions in the space of orbits. These four classes of orbits also could not be distinguished at the level of dynamics in flat space-time, as all of them  describe  massless particles (with only one independent momentum vector).  Surprisingly, the  orbits $iic)$ also describe  massless particles, despite  their  invariants  $ {\cal C}^{(2)}$  and $ {\cal C}^{(4)}$ taking the same values as those of   massive particles  $ia)$.  Unlike with  $ib)$, $iib)$, $iii)$ and $iv)$,  the massless particles  $iic)$ possess extra momentum degrees of freedom, as $\tilde p^a$ is not fully determined from  $p^a$.  The  physical meaning of these  extra degrees of freedom is not clear and worth further investigation.  Moreover, the list of orbits given in the table at the end of section four is by no means complete.   A more complete classification could be of interest - in particular - with regards to the pursuit of dark matter candidates.
 
In section 5.3 we obtained the general solutions to the equations of motion (for particles with no spin or charge) in an arbitrary background, which is characterized  by  $e^a_{\mu}$ and  $f^a_{\mu}$. For orbits $ia)$ and $iic)$ we obtained an effective Lagrangian (\ref{efctvlgrngn}) which contained corrections to the naive Lagrangian (\ref{mstobvsS}).  The corrections vanished in the limit where the  vierbein fields $f^a_\mu$ vanish, and so one recovers geodesic motion in this limit.   On the other hand, corrections to geodesic motion do occur in the more general setting.  Therefore deviations from geodesic motion, such as the  reported  Pioneer anomaly\cite{Anderson:1998jd}, could signal the presence of additional fields in gravity like the vierbeins   $f^a_\mu$.

The particle spin, along with two $U(1)$ charges were taken into account in section   six.   There we found three more  independent $GL(2,C)$ invariants ${\cal C}^{(2)}_w$, ${\cal C}^{(4)}_w$ and ${\cal C}^{(2)}_{p,w}$ in (\ref{wwtldcsmr}) and (\ref{nvrntwthpnw}), which were constructed using two Pauli-Lubanski vectors (\ref{twoplvctrs}).   It then follows that at least six invariants   (not including the two charges) are needed to classify spinning particles in this theory -  as opposed to the usual two, i.e., mass and the square of the Pauli-Lubanski vector.  The dynamical equations for this system, including the interactions with the two $U(1)$ fields and two sets of torsion tensors, were obtained  using the Wess-Zumino term for $GL(2,C)$.
Their solutions should lead to further deviations   from geodesic motion.

We wrote down the algebra of quantum mechanical observables in section eight, and showed that the usual method of induced representations can be applied to construct the Hilbert space.  It remains to develop the  $n$-particle interacting system and also the  field theory associated with the various particle representations.  (The coupling to fermions and generalization to supergravity was recently studied in \cite{Aschieri:2009ky}.)

Since the  $GL(2,C)$ gauge theory has the advantage of being amenable to a  noncommutative generalization,  it is then natural to also promote the particle dynamics to the  noncommutative setting. One approach would be to take the infinite field limit of one of the $U(1)$ gauge fields, as nontrivial space-time commutation relations  result from canonical quantization.\footnote{I thank A. Pinzul for this remark.}  Alternatively,  or in addition, one can search for solutions to the noncommutative field equations, and  
then apply a Seiberg-Witten map back to the commutative theory.  One can thereby obtain corrections in  the metric tensor (\ref{gltcmtrc}), as well as other $GL(2,C)$ invariant quantities, for known gravity solutions, such as 
 black hole  and cosmological solutions.\footnote{This differs from previous  approaches to finding noncommutative corrections to the solutions of general relativity\cite{Pinzul:2005ta},\cite{Balachandran:2006qg},\cite{Dolan:2006hv},
\cite{Chaichian:2007we},\cite{Mukherjee:2007fa},\cite{Kobakhidze:2007jn},\cite{Banerjee:2008gc},\cite{Buric:2008th},\cite{Nicolini:2008aj},\cite{Fabi:2008ta}.  For example, in some of those works a  noncommutative analogue of the metric tensor had to be defined in order to make a physical interpretation.
  On the other hand, the approach here utilizes the commutative metric tensor (\ref{gltcmtrc}).
What we are proposing  is similar in spirit to \cite{Stern:2008wi}, where,  in the context of Maxwell theory, noncommutative corrections were found to the Coulumb solution.  It is also similar to \cite{Marculescu:2006ca}, where first order noncommutative corrections were found for $pp-$wave solutions to the coupled Einstein-Maxwell equations.} In this case, the  fields degrees of freedom  $a_\mu$, $b_\mu$ and $f^a_\mu$ associated with the $GL(2,C)$ central extension, which are zero for the familiar gravity solutions, will in general pick up  nonvanishing contributions after applying the Seiberg-Witten map from the noncommutative solutions.  Moreover, it is expected that these contributions are  first order in the noncommutativity parameter.  First order corrections to geodesic motion may then result, and are computable using the results found here.

\end{document}